\documentclass[aps,pra,reprint]{revtex4-2}
\usepackage{amsmath}
\usepackage{amssymb}
\usepackage{graphicx}
\usepackage{mathrsfs}
\usepackage{hyperref}
\usepackage{braket}
\usepackage{todonotes}
\usepackage{booktabs}
\usepackage[caption=false]{subfig}
\hypersetup{colorlinks,linkcolor=blue,citecolor=blue,urlcolor=blue}

\graphicspath{{QKNH2_figures/}}

\begin{document}


\title{Quantum dynamical transitions when the corresponding classical phase space has a separatrix: extension of the QKNH theorem beyond double wells}

\author{Peter Stabel}
\author{James R. Anglin}

\affiliation{\mbox{State Research Center OPTIMAS and Fachbereich Physik,} \mbox{RPTU Kaiserslautern-Landau,} \mbox{D-67663 Kaiserslautern, Germany}}

\date{\today}

\begin{abstract}

In classical Hamiltonian systems with a slowly time-dependent parameter, adiabaticity breaks down at a separatrix, leading to probabilistic post-adiabatic transitions between different phase-space regions described by the Kruskal-Neishtadt-Henrard (KNH) theorem. Quantum mechanically, dynamical tunneling allows adiabaticity to persist where it breaks down classically, making the quantum-classical correspondence for separatrix crossing non-trivial. We recently established a quantum version of the KNH theorem (QKNH) by introducing the concept of a sweep-rate-dependent quantum separatrix in the instantaneous energy spectrum of the time-dependent system. The QKNH theorem relates the splitting of a quantum ensemble between the outgoing dynamical sectors at the quantum separatrix to their relative growth rates, and thus provides a quantum analog of the classical KNH rule. Here, we demonstrate that these findings are not restricted to double-well systems with tunneling through a potential barrier. We investigate a slowly driven quantum pendulum, in which the avoided crossings arise from above-barrier reflection. The resulting Landau-Zener probabilities have the same local lattice form as in the double-well problem, allowing the application of the QKNH theorem. 

\end{abstract}
\maketitle

\section{Introduction}
\label{sec:intro}
\subsection{Kruskal-Neishtadt-Henrard Theorems}
\subsubsection{The classical KNH theorem}
The Kruskal-Neishtadt-Henrard (KNH) theorem of classical mechanics \cite{NEISHTADT1975594,Neishtadt_2017,henrard_capture_1982,LiouvilleControl} governs separatrix-crossing dynamical transitions between disjoint phase space regions, when a system evolves under a slowly time-dependent Hamiltonian. The different phase space regions between which the system moves are usually distinguished by qualitatively different forms of time evolution for the system, and so crossing a separatrix is a decisive event that can change the type of classical motion, for example from rotation to libration. It can be difficult to predict, however, because a separatrix includes an unstable fixed point, and so these transitions involve breakdown of adiabaticity and sensitive dependence on initial conditions. The KNH theorem derives conditions and probabilities for separatrix-crossing transitions, based on Liouville's theorem, in terms of the relative rates at which the phase space regions that are separated by the separatrix grow or shrink as the Hamiltonian changes.
\subsubsection{The quantum KNH theorem}
How does the classical KNH theorem extend to quantum mechanics? The question is subtle, because the classical and adiabatic limits do not in general commute \cite{Berry_1984}. We recently extended the KNH theorem into the semi-classical WKBJ limit of quantum mechanics \cite{Stabel_2022} by analyzing Landau-Zener transitions in a network of avoided crossings in the adiabatic spectrum of a quantum double-well model. Our extension begins by noticing that tunneling through the potential barrier effectively creates a quantum separatrix which is shifted away from the classical separatrix when the Hamiltonian change rate is slow. Once this rate-dependent quantum separatrix is correctly defined and the corresponding growth rates of different dynamical quantum sectors are computed accordingly, a Quantum Kruskal-Neishtadt-Henrard theorem (QKNH) follows from unitarity, as the quantum analog of Liouville's theorem. 
\subsection{Tunneling and the Quantum Separatrix}

\subsubsection{Quantum separatrix in a double well}
 The QKNH clarification of quantum-classical correspondence in separatrix crossing came through a semi-classical treatment of tunneling. In a lattice of many avoided crossings of energy levels as functions of a control parameter, the crossings are narrowly avoided because degeneracies between energy eigenstates localized in each of the two wells are lifted by tunneling. So the narrow energy gaps in these crossings are proportional, in the semi-classical limit, to an exponential factor $e^{-T_b/\hbar}$, where $T_b$ is the imaginary part of a classical action in an analytically continued phase space that includes imaginary momentum in the classically forbidden region under the potential barrier. 
 
 Since semi-classical energy levels differ by order $\hbar$, the ratios of the gaps $\Delta_2/\Delta_1 \simeq e^{(T_{b}^{(1)}-T_b^{(2)})/\hbar}$ among nearby avoided crossings (labeled 1,2), are of order $\hbar^0$ in the semi-classical limit, and depend only on classical actions (albeit analytically continued). Because Landau-Zener transition probabilities depend exponentially on the squared gaps, the avoided crossings can change within just a few energy levels from almost completely diabatic to almost completely adiabatic. 
 
 This abrupt change in quantum time evolution across a narrow range of energies is in close analogy to the qualitative change in classical dynamics across a classical separatrix. The major difference is that the energy of the quantum separatrix depends in general on the rate at which the Hamiltonian changes, and can thereby differ from the classical separatrix energy.

\subsubsection{Beyond the semiclassical double well}
In \cite{Stabel_2022} we explicitly studied only double-well systems in the semi-classical limit, but we suggested that pendulum-like separatrices should also have quantum analogs, and obey the same form of QKNH theorem. We also proposed that the concept of a quantum separatrix zone---a narrow energy range, in a network of avoided crossings, across which the probability of a crossing being non-adiabatic should change from nearly zero to nearly one---could extend beyond the semi-classical regime, and that evolution through such a quantum separatrix zone would obey the QKNH theorem even if the semi-classical limit did not apply. 

Important aspects of these conjectures have been confirmed by Quillen, Skerrett, Sowinski and Miakhel \cite{harpermodel}, in a study of quantum dynamical transitions under slowly varying parameters in the quantized Harper model. This model is a pendulum-like system having a finite-dimensional Hilbert space because the corresponding classical phase space is a finite torus. Reference \cite{harpermodel} does not rely on semi-classical analysis, but instead performs direct numerical quantum evolution through the unitary propagator and evaluates transition probabilities between initial and final instantaneous eigenstates. Quillen \emph{et al.} sum these transition probabilities over regions corresponding to classical libration and circulation. Averaging these transition probabilities between dynamical sectors over the initial states corresponds to an equally weighted initial ensemble, directly connecting their calculation to our ensemble formulation of the QKNH theorem in \cite{Stabel_2022}. They identify a regime of sufficiently slow sweep rate in which transitions between instantaneous eigenstates are suppressed in the librating regions, while diabatic transitions persist in the circulating region. Such diabatic passage through narrow avoided crossings is consistent with preserving the classical circulation branch. They use the classical KNH rule for a pendulum approximation of the Harper model to estimate capture from regions corresponding to classical circulation into regions corresponding to libration and compare these predictions with the transition probabilities between dynamical sectors in the quantum process. In particular, for sufficiently slow sweeps, the transfer probabilities between dynamical regions become nearly independent of the sweep rate. These results provide support for the quantum KNH picture in a complementary finite-dimensional system.
\subsubsection{The semiclassical pendulum}
The further extension of QKNH theory beyond semi-classical analysis, as begun in \cite{harpermodel}, is an important goal for future work. In the present paper, however, we return to the semi-classical regime of \cite{Stabel_2022}, to confirm our conjectures about QKNH pendulum-like separatrices semi-classically. Although we will focus explicitly on the particular model of a simple physical pendulum, the applicability of semi-classical methods to a wide range of dynamical systems makes the implications of our results quite general. Our previous paper addressed the case where three phase space regions are separated by a separatrix and two regions have lower energy while the third is higher; we now supply the complementary case where two regions have higher energy and the third is lower. Our preceding work considered the case where the quantum separatrix can diverge from the classical one due to tunneling through a potential barrier; we now supply the complementary case where the tunneling occurs in momentum-space, providing above-barrier reflection \cite{Jaffe, Heller1999}. Combining both cases shows that what can move the quantum separatrix away from the classical one, and what makes QKNH different in general from KNH, is not specifically tunneling in position, through barriers, but more generally, \emph{dynamical tunneling}.
\subsection{Dynamical transitions for a classical pendulum}
\subsubsection{Classical pendulum model}
\label{subsec:pendulum_KNH}
\begin{figure}
    \centering
    \includegraphics[width=\linewidth]{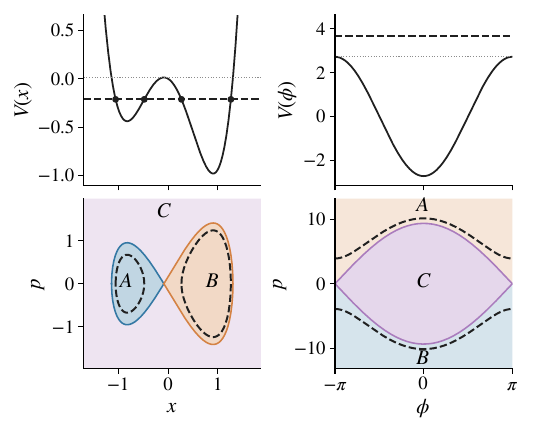}
    \caption{Potential and corresponding phase space picture for an asymmetrical double well system (left) and a pendulum system (right). For suitable energies below the central barrier, there exist two energy-degenerate phase space contours. The pendulum instead has two such energy-degenerate rotational contours above the barrier.}
    \label{fig:intro}
\end{figure}

We consider the Hamiltonian
\begin{equation}
H(p,\phi;\lambda) =  \left[\frac{\Big(p-\alpha(\lambda)\Big)^2}{2\mu} - \beta^2(\lambda)\cos\phi\right]\;.
\label{eq:pendulum_hamiltoanian}
\end{equation}
Here, $\lambda$ is an external sweep parameter. The classical momentum $p$ and shift-parameter $\alpha$ have units of angular momentum and $\mu$ has units of a moment of inertia. This Hamiltonian is a standard model for resonance capture, in which slow parameter variation drives a separatrix crossing and captures an initially unbound particle into bound motion \cite{henrard_capture_1982,Fieguth_generalized_KNH}. At fixed $\mu$, Hamilton's equations give $\mu\ddot{\phi}=-\beta^2\sin\phi-\dot{\alpha}$, so $-\dot{\alpha}$ acts as an applied torque which is constant for the linear sweep protocol, while a variation of $\beta^2$ can be associated with a change in pendulum length \cite{henrard_capture_1982}. In this article, we assume $\mu, \beta>0$, and we consider linear parameter sweeps
\begin{align}
\alpha(\lambda) &= \alpha_i + (\alpha_f-\alpha_i)\lambda,\label{alphasweep}\\
\beta(\lambda) &= \beta_i + (\beta_f-\beta_i)\lambda,
\label{betasweep}
\end{align}
\label{sec:KNH and QKNH}
where the dimensionless sweep parameter is defined by
\begin{equation}
\lambda(t)=\frac{t}{T}\in[0,1]
\end{equation}
and T is the duration of the parameter sweep. Changing $\alpha$ translates the separatrix in momentum, while increasing $\beta$ increases the phase-space area enclosed by the separatrix. The pendulum phase space consists of three dynamically distinct regions labeled A, B and C in Fig.~\ref{fig:intro}, separated by an eye-shaped separatrix. The separatrix energy $E_{s} = \beta^2$ coincides with the height of the cosine barrier. We restrict our attention to protocols for which region A shrinks and both B and C grow. For $\alpha',\beta'>0$ this requires $\pi\alpha'>4\sqrt\mu\,\beta'$. At fixed energy, the two momentum branches are 
\begin{equation}
    p_\pm(\phi,E,\lambda) = \alpha\pm\sqrt{2\mu(E+\beta^2(\lambda)\cos\phi)}. 
\end{equation}
For $E>\beta^2$, these define two separate energy-degenerate rotational orbits corresponding to counter-clockwise (region A) and clockwise (region B) rotational motion of the pendulum, whereas below the separatrix, they combine to form librational orbits (region C), with a single non-degenerate contour for each energy.
\subsubsection{Classical KNH transition probabilities}
We consider a trajectory contained initially in region A with initial action $I_0$ which is approximately conserved adiabatically. Note that in the pendulum case, the initial action is identified with the cut-off referenced invariant $I_0=\Sigma_A(E_0,\lambda_0)/(2\pi)$ and $\Sigma_A$ is defined by eq.~\eqref{Sigma_A}. Hence, the trajectory encounters the separatrix when the separatrix lobe (in the double well case) or the rotational phase space region (in the pendulum case) the trajectory is in shrinks to the corresponding area $\mathcal{A}_A(\lambda_*) = 2\pi I_0$ where $\lambda_*=\lambda(t_*)$ is the sweep parameter value at the time of the encounter. Every trajectory initially contained in region A that encounters the separatrix subsequently leaves region A and enters either region B or region C. The Kruskal-Neishtadt-Henrard theorem states that the corresponding transition probabilities are given as the ratios of the relevant area growth rates \cite{NEISHTADT1975594,henrard_capture_1982,LiouvilleControl}: 
\begin{equation}
P_{A\rightarrow B,C}=-\frac{\displaystyle \frac{d}{d\lambda}\mathcal{A}_{B,C}(\lambda)}
{\displaystyle \frac{d}{d\lambda}\mathcal{A}_A(\lambda)},
\end{equation}
where $\mathcal{A}_{A,B,C}$ denote the phase-space areas of the three regions separated by the separatrix. These fractions are to be evaluated at the parameter value $\lambda_*$ of the trajectory's encounter with the separatrix. 
For energies above the separatrix energy, we can associate the following areas with the rotational orbits: 
\begin{align}
\label{Sigma_A}
&\Sigma_A(E,\lambda)
=
\int_{-\pi}^{\pi}
[p_{\mathrm u}-p_+(\phi,E,\lambda)]\,d\phi,\\
&\Sigma_B(E,\lambda)
=
\int_{-\pi}^{\pi}
[p_-(\phi,E,\lambda)-p_{\mathrm l}]\,d\phi.
\end{align}
Here $p_{u,l}$ are fixed upper and lower momentum cutoffs chosen outside the orbits under consideration. Then the areas of the full rotational regions are these areas evaluated at the separatrix energy: $\mathcal{A}_{A,B} = \Sigma_{A,B}(E_s(\lambda),\lambda)$, and the area enclosed by the separatrix is 
\begin{equation}
\mathcal{A}_C(\lambda)
=
\int_{-\pi}^{\pi}
\left[
p_+(\phi,E_s(\lambda),\lambda)
-
p_-(\phi,E_s(\lambda),\lambda)
\right]d\phi.
\end{equation}
These evaluate to (see also \cite{LiouvilleControl})
\begin{align}
    &\mathcal{A}_C = 16\beta\sqrt{\mu}\\
    &\mathcal{A}_A = C_A-2\pi\alpha-\frac{A_C}{2}\\
    &\mathcal{A}_B = C_B+2\pi\alpha-\frac{A_C}{2} 
\end{align}
with the constant reference areas $C_A=2\pi p_u$ and $C_B = -2\pi p_l$. For the pendulum Hamiltonian, the transition probability from region A into the librational region C or the other rotational region $B$ is then given by 
\begin{align}
&P_{A\to C} = -\frac{\dot{\mathcal{A}}_C}{\dot{\mathcal{A}}_A}=\frac{8\sqrt{\mu}\,\beta'}
{\pi\alpha'+4\sqrt{\mu}\,\beta'}\\\label{KNH_AB}
&P_{A\to B}=-\frac{\dot{\mathcal{A}}_B}{\dot{\mathcal{A}}_A}=\frac{\pi\alpha'-4\sqrt{\mu}\,\beta'}
{\pi\alpha'+4\sqrt{\mu}\,\beta'}=1-P_{A\to C}.
\end{align}
where the primes denote derivatives with respect to $\lambda$. For the protocols considered here, these fractions are independent of the encounter point since $\alpha,\beta$ are linear functions of $\lambda$, so the area growth rates are constant.

\section{Quantum KNH theorem applied to the pendulum model}
In the following, we show that the semi-classical construction of the avoided-crossing network and resulting Landau-Zener probabilities derived in \cite{Stabel_2022} can analogously be obtained for the quantum pendulum, demonstrating that the quantum KNH treatment is not restricted to cases where gaps open due to tunneling through a physical barrier but also applies when gaps open due to above-barrier reflection. 
\label{subsec:QKNH}
\begin{figure}
    \centering
    \includegraphics[width=\linewidth]{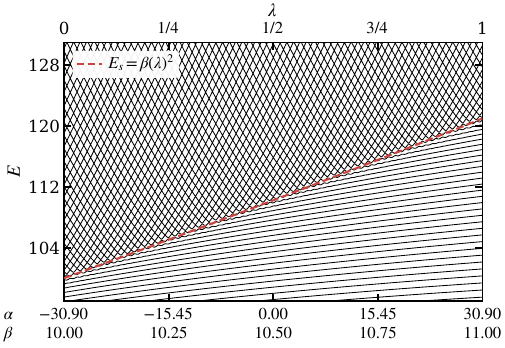}
    \caption{Instantaneous eigenenergies of the quantum pendulum system under a parameter sweep protocol, with $\mu=50$ and $\hbar =1$. The red curve indicates the classical separatrix energy $E_{s}(\lambda)=\beta^2(\lambda)$ which separates rotational and librational motion in the classical pendulum.}
    \label{fig:spectrum}
\end{figure} 
\subsection{Instantaneous spectrum of the quantum pendulum}
The classical Hamiltonian is quantized by the usual procedure of promoting the canonical variables to hermitian operators with the usual representation of the momentum operator $\hat{p}=-i\hbar\partial_\phi$ in the coordinate basis. The resulting dynamics are governed by the time-dependent Schrödinger equation
\begin{equation}
i\hbar\frac{\partial}{\partial t}\ket{\psi(t)}
=
\hat{H}(\alpha(\lambda(t)),\beta(\lambda(t)))
\ket{\psi(t)},
\end{equation}
where the time dependence enters through the parameter sweep protocol \eqref{alphasweep}, \eqref{betasweep}. The corresponding instantaneous eigenstates satisfy
\begin{equation}
\hat{H}(\lambda)\ket{E_j(\lambda)}
=
E_j(\lambda)\ket{E_j(\lambda)}.
\end{equation}
Fig.~\ref{fig:spectrum} shows the instantaneous spectrum over sweep parameter $\lambda$. The classical phase-space structure shown in Fig.~\ref{fig:intro} is reflected in the quantum spectrum. Below the classical separatrix energy the instantaneous eigenstates form a single ladder associated with librational motion whereas above it the spectrum exhibits a network of avoided crossings.
\subsection{Semiclassical quantization}
To explain this structure and obtain the transition probabilities at the avoided crossings, we use the pendulum's semiclassical quantization condition. For the evaluation of the relevant action integrals, it is convenient to use the shifted angle $\theta = \phi-\pi$ which shifts the maximum of the cosine barrier to $\theta=0$ without changing the actions. The quantum energies are restricted by the semiclassical quantization condition which follows from WKBJ connection formulas and periodic boundary conditions on the wave function and its derivative (see appendix \ref{app:quantcond}):
\begin{equation}
\label{quandcond}
\sqrt{1+e^{-2R_B/\hbar}}\cos\!\left(
\frac{\tilde S_A+\tilde S_B}{\hbar}
\right)
=
\cos\!\left(
\frac{\tilde S_A-\tilde S_B}{\hbar}
\right)
\end{equation}
where $\tilde{S}_{A,B}$ are signed phase-corrected half areas associated with the rotational orbits with the energy-independent cutoffs removed, and $R_b$ is the integral associated with above-barrier reflection. The signed half-areas are 
\begin{align}
\label{S_AB}
S_A
&=\frac{\Sigma_A(E,\lambda)-C_A}{2}
=-\frac{S}{2}-\pi\alpha,\\
S_B
&=\frac{\Sigma_B(E,\lambda)-C_B}{2}
=-\frac{S}{2}+\pi\alpha,
\end{align}
and
\begin{align}
\label{S}
S(E,\lambda)
&=
\int_{-\pi}^{\pi}
\sqrt{2\mu(E-\beta^2\cos\theta)}\,d\theta\nonumber\\
&= 8\beta\sqrt{\mu}\sqrt{\kappa}\mathrm{E}(\kappa^{-1})
\end{align}
with $\kappa=(E+\beta^2)/(2\beta^2)$ and $\mathrm{E}$ is the complete elliptic integral of the second kind (see appendix \ref{app:elliptic_integrals}). Note that evaluated at the separatrix energy, this is the half area enclosed by the eye-shaped classical separatrix, \textit{i.e.} $S(E_s=\beta^2,\lambda) = \mathcal{A}_C(\lambda)/2$. The tilde indicates a phase correction \cite{CHILD1974280} from the barrier connection formula:
\begin{align}
\label{tilde_SAB}
    &\tilde{S}_{A,B} = S_{A,B}+\frac{\hbar\Phi}{2}=-\frac{\tilde{S}}{2}\mp\pi\alpha\\
    &\tilde{S} = S-\hbar\Phi\\
    &\Phi = \arg\Gamma\left(\frac{1}{2}+\frac{iR_b}{\pi\hbar}\right) - \frac{R_b}{\pi\hbar}\left(\log\left|\frac{R_b}{\pi\hbar}\right|-1\right)\label{Phase_correction}
\end{align}
The gamma-function phase is chosen continuously for $x=R_b/(\pi\hbar)>0$, with $\arg\Gamma(1/2)=0$. Equivalently, we use the continuous log-gamma function in eq.~\eqref{Phi_Imlog}. Although the rotational motion has no real turning points above the barrier, above barrier reflection couples the two rotational branches. The relevant complex turning points are $\theta_\pm = \pm i\eta$ with $\eta=\operatorname{arcosh}(E/\beta^2)$. The above-barrier reflection integral is defined as \cite{CHILD1974280}:
\begin{align}
\nonumber
    R_b &= \left|\mathrm{Im}\int_{\theta_-}^{\theta_+}\mathrm{d}\theta\;\sqrt{2\mu(E-\beta^2\cos\theta)}\right|\\&=\int_{-\eta}^{\eta}\mathrm{d}y \sqrt{2\mu(E-\beta^2\cosh y)}\nonumber\\
    &=8\beta\sqrt{\mu}\left[\kappa\mathrm{K}(1-\kappa)-\mathrm{E}(1-\kappa)\right]\label{R}
\end{align}
where $\mathrm{K}$ is the complete elliptic integral of the first kind and $R_b>0$. The quantization condition can be equivalently written as 
\begin{align}
\label{BS_new}
    \tan\left(\frac{\tilde{S}_A}{\hbar}\right)\tan\left(\frac{\tilde{S}_B}{\hbar}\right) &=  \left[\frac{\sqrt{1+e^{-2R_b/\hbar}}-1}{\sqrt{1+e^{-2R_b/\hbar}}+1}\right]\\
    &= \frac{e^{-2R_b/\hbar}}{4} + \mathcal{O}(e^{-4R_b/\hbar})
\end{align}
provided $\cos(\tilde{S}_A/\hbar)\cos(\tilde{S}_B/\hbar)\neq0$. In the weak reflection limit, $R_b/\hbar\gg1$, so $\varepsilon = e^{-R_b/\hbar}\ll1$. In zeroth order in $e^{-2R_b/\hbar}$ this yields the two uncoupled rotational ladders 
\begin{equation}
\label{zero_order_quantcond}
\tilde{S}_A(E^A_m(\lambda),\lambda) =\pi\hbar m,\qquad \tilde{S}_B(E^B_n(\lambda),\lambda) =\pi\hbar n
\end{equation}
for integers $m,n$. The intersections of these ladders satisfy $E^A_m(\lambda_{mn})=E^B_n(\lambda_{mn})=E_{mn}$. Subtracting the two uncoupled quantization conditions at such an intersection eliminates the common energy-dependent contribution and gives $2\alpha/\hbar=n-m\in\mathbb Z$. The rotational ladders therefore intersect at integer multiples of $\hbar/2$ in $\alpha$, and above-barrier reflection opens gaps at these intersections, turning them into avoided crossings. 
\subsection{Local Mapping to an effective Landau-Zener Hamiltonian and transition probabilities}
We introduce the energy and sweep parameter displacements
$\delta E=E-E_{mn}$,\ $\delta\lambda=\lambda-\lambda_{mn}$ and define $\varepsilon=e^{-R_b(E_{mn},\lambda_{mn})/\hbar}$. We assume $\delta E = \varepsilon e$, $\delta\lambda=\varepsilon\ell$ with $e,\ell$ independent of $\varepsilon$. Expanding the half-areas near the intersection gives
\begin{align}
\label{SA_expansion}
&\tilde S_A(E,\lambda)
=
\pi\hbar m
+\partial_E\tilde S_A\,\delta E
+\partial_\lambda\tilde S_A\,\delta\lambda
+\cdots,\\\label{SB_expansion}
&\tilde S_B(E,\lambda)
=
\pi\hbar n
+\partial_E\tilde S_B\,\delta E
+\partial_\lambda\tilde S_B\,\delta\lambda
+\cdots,
\end{align}
where all derivatives are to be evaluated at $(E,\lambda) = (E_{mn},\lambda_{mn})$. Since the parameter dependence enters through $\alpha(\lambda)$ and $\beta(\lambda)$, differentiation at fixed energy gives
\begin{align}
    \frac{\partial \tilde{S}_{A,B}}{\partial\lambda}&=\frac{\partial \tilde{S}_{A,B}}{\partial\alpha}\frac{\partial\alpha}{\partial\lambda} + \frac{\partial \tilde{S}_{A,B}}{\partial\beta}\frac{\partial\beta}{\partial\lambda}\\
    &=\mp\pi\frac{\partial\alpha}{\partial\lambda}-\frac{1}{2}\frac{\partial\tilde{S}}{\partial\beta}\frac{\partial\beta}{\partial\lambda}.
\end{align}
Insertion of the expansions \eqref{SA_expansion}, \eqref{SB_expansion} into the quantization condition \eqref{quandcond} and retaining the leading $\mathcal{O}(\varepsilon^2)$ terms yields
\begin{align}
\label{expanded_quantcond}
&\left(
\partial_E\tilde S_A\,\delta E
+\partial_\lambda\tilde S_A\,\delta\lambda
\right)
\left(
\partial_E\tilde S_B\,\delta E
+\partial_\lambda\tilde S_B\,\delta\lambda
\right)\nonumber\\
&=\frac{\hbar^2}{4}e^{-2R_b/\hbar}.
\end{align}
In zeroth order in $\varepsilon$, the energies $E_m^A, E_n^B$ are defined by the quantization rules \eqref{zero_order_quantcond}. Differentiating them with respect to $\lambda$ results in 
\begin{align}
&\frac{d\tilde{S}_A}{d\lambda} = \frac{\partial \tilde{S}_A}{\partial\lambda} + \frac{\partial \tilde{S}_A}{\partial E}\frac{dE}{d\lambda} = \frac{d}{d\lambda}(\pi\hbar m) = 0\nonumber\\
\label{dEdlambda}
&\Leftrightarrow \frac{dE^A_m}{d\lambda} = -\frac{\partial_{\lambda}\tilde{S}_A}{\partial_E \tilde{S}_A}
\end{align}
and analogously for $E_B^n$. Dividing \eqref{expanded_quantcond} by $\partial_E\tilde S_A\,\partial_E\tilde S_B$ yields 
\begin{equation}
\left(
\delta E-\frac{dE_m^A}{d\lambda}\delta\lambda
\right)
\left(
\delta E-\frac{dE_n^B}{d\lambda}\delta\lambda
\right)
-
\frac{\hbar^2e^{-2R_b/\hbar}}
{4\partial_E\tilde S_A\,\partial_E\tilde S_B}
=0
\end{equation}
which is the characteristic equation $\det(\hat{H}_\mathrm{eff}-E\mathbf{1})=0$ of the following local two-state Hamiltonian in the diabatic basis of states $\ket{E_m^A}, \ket{E_n^B}$ and eigenvalue $E=E_{mn}+\delta E$:
\begin{align}
&\hat H_{\mathrm{eff}}
=
E_{mn}\mathbf1+
\begin{pmatrix}
\dfrac{dE_m^A}{d\lambda}\delta\lambda&V_{mn}\\[4pt]
V_{mn}&\dfrac{dE_n^B}{d\lambda}\delta\lambda
\end{pmatrix},
\\
&V_{mn}
=
\frac{\hbar e^{-R_b/\hbar}}
{2\sqrt{\partial_E\tilde S_A\,
              \partial_E\tilde S_B}}.
\end{align}
The minimum gap separating the two levels that meet at the avoided crossing is $2V_{mn}$, and neglecting the phase correction yields the established semiclassical rotational splitting for the pendulum \cite{Vybornyi2014}: 
\begin{equation}
\Delta E_{\min}
= \frac{\hbar\omega_{\mathrm{rot}}}{\pi}e^{-R_b/\hbar},\qquad\omega_{\mathrm{rot}}=2\pi/{{\mathrm{\partial_E S}}}
\end{equation}
where $\omega_\mathrm{rot}$ is the frequency associated with the classical period $T_c = \partial_E S$. For a locally linear sweep $\delta\lambda = \dot{\lambda}(t_{mn})(t-t_{mn})$, the local Hamiltonian takes the Landau-Zener form
\begin{align}
\hat H_{\mathrm{eff}}(t)
=
&\bar E(t)\mathbf1+
\frac{\dot\lambda(t_{mn})}{2}
\left(
\frac{dE_m^A}{d\lambda}
-\frac{dE_n^B}{d\lambda}
\right)(t-t_{mn})\sigma_z
\nonumber\\&+V_{mn}\sigma_x,
\end{align}
where
\begin{equation}
\bar E(t)
=
E_{mn}
+\frac{\dot\lambda(t_{mn})}{2}
\left(
\frac{dE_m^A}{d\lambda}
+\frac{dE_n^B}{d\lambda}
\right)(t-t_{mn}),
\end{equation}
and $\sigma_x = \ket{E_n^B}\bra{E_m^A} + \ket{E_m^A}\bra{E_n^B}$, $\sigma_z = \ket{E_m^A}\bra{E_m^A}-\ket{E_n^B}\bra{E_n^B}$ are the Pauli matrices. The first term contributes only a common phase. Assuming an isolated crossing, comparison with the Landau-Zener Hamiltonian $H_LZ = ut\sigma_z + V\sigma_x$ yields the diabatic survival probability $P_{LZ}=e^{-\pi V^2/(\hbar|u|)}$:
\begin{equation}
\label{pmn}
P_{mn}
=
\exp\!\left[
-\frac{2\pi V_{mn}^2}
{\hbar|\dot\lambda|\,
\left|
\dfrac{dE_m^A}{d\lambda}
-\dfrac{dE_n^B}{d\lambda}
\right|}
\right].
\end{equation}

Using eq. \eqref{dEdlambda} gives
\begin{equation}
    \partial_E\tilde S_A\,
\partial_E\tilde S_B
\left(
\frac{dE_m^A}{d\lambda}
-\frac{dE_n^B}{d\lambda}
\right)
=
[\tilde S_A,\tilde S_B]_{E,\lambda},
\end{equation}
where
\begin{equation}
    [F,G]_{E,\lambda}
=
\partial_EF\,\partial_\lambda G
-
\partial_\lambda F\,\partial_EG.
\end{equation}
Substitution into \eqref{pmn} finally yields  
\begin{align}\label{plz}
P_{mn}
&=
\exp\!\left[
-\frac{\pi\hbar e^{-2R_b/\hbar}}
{2|\dot\lambda|\,
|[\tilde S_A,\tilde S_B]_{E,\lambda}|}
\right]_{(E,\lambda)=(E_{mn},\lambda_{mn})}
\end{align}
\begin{figure}[t]
    \centering
    \includegraphics[width=0.49\textwidth]{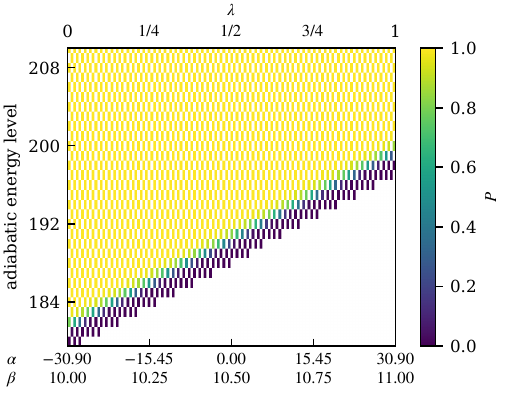}%
    \caption{Heatmap of the Landau-Zener probabilities given by \eqref{plz}: each colored cell in this figure represents an avoided crossing in the instantaneous spectrum of the pendulum in Fig. \ref{fig:spectrum} under the parameter sweep used in that figure with a sweep time $T = 10^5$. The color code shows the Landau-Zener transition probability $P_{j\rightarrow j+1}$ = $P_{j+1\rightarrow j}$ between adiabatic levels $E_{j}, {E_{j+1}}$ that meet at the avoided crossings. The phase correction of the areas is retained here. } 
    \label{fig:LZ_heatmap}   
\end{figure}
Figure ~\ref{fig:LZ_heatmap} shows the resulting transition probabilities at the avoided crossings for an illustrative sweep protocol. Away from the transition region which constitutes the quantum separatrix zone, the avoided crossings are either almost perfectly diabatic ($P_{LZ}\approx1$) or almost perfectly adiabatic ($P_{LZ}\approx0$). Only a narrow band of crossings contributes significantly to the redistribution of probability.  
\subsection{Lattice of avoided crossings}
Under the assumption that there is a dense lattice of avoided crossings, we can construct this lattice of avoided crossings as in the case of the double well model treated in Ref. \cite{Stabel_2022}. We choose the reference intersection point $(E_{m_0n_0},\lambda_{m_0n_0})$ of the uncoupled ladders, and use the simple Bohr-Sommerfeld rules \eqref{zero_order_quantcond} to linearize the phase corrected half-areas around this reference point. This yields the values at another nearby intersection ($E_{mn},\lambda_{mn})$:
\begin{align}
\tilde{S}_A(E_{mn},\lambda_{mn}) \approx & \tilde{S}_A(E_{m_0n_0},\lambda_{m_0n_0}) + \partial_E\tilde{S}_A(E_{mn}-E_{m_0n_0})\nonumber  \\&+\partial_\lambda \tilde{S}_A(\lambda_{mn}-\lambda_{m_0n_0}) =\pi\hbar m
\end{align}
All partial derivatives are to be evaluated at $(E_{m_0n_0},\lambda_{m_0n_0})$. Writing the analogous expression for $\tilde{S}_B$ yields
\begin{equation}
\begin{pmatrix}
\partial_E\tilde S_A&\partial_\lambda\tilde S_A\\
\partial_E\tilde S_B&\partial_\lambda\tilde S_B
\end{pmatrix}
\begin{pmatrix}
E_{mn}-E_{m_0n_0}\\
\lambda_{mn}-\lambda_{m_0n_0}
\end{pmatrix}
=
\pi\hbar
\begin{pmatrix}
m-m_0\\
n-n_0
\end{pmatrix}.
\end{equation}
Inversion of this matrix relation defines a lattice of avoided crossings in the $(E,\lambda)$ plane:
\begin{align}
&\delta E_{mn}
=
\frac{\pi\hbar}{
[\tilde S_A,\tilde S_B]}
\left(
\delta m\,\partial_\lambda\tilde S_B
-\delta n\,\partial_\lambda\tilde S_A
\right)\\\label{lambda_mn}
&\delta\lambda_{mn}
=
\frac{\pi\hbar}{
[\tilde S_A,\tilde S_B]}
\left(
\delta n\,\partial_E\tilde S_A
-\delta m\,\partial_E\tilde S_B
\right).
\end{align}
where $\delta E_{mn} = E_{mn}-E_{m_0,n_0}$, $\delta\lambda_{mn} = \lambda_{mn}-\lambda_{m_0,n_0}$, $\delta m = m-m_0$, $\delta n = n-n_0$. In addition, we can expand the reflection integral $R_B$ in the vicinity of the reference crossing:
\begin{align}
    R_b(E_{mn},\lambda_{mn}) \approx & R_b(E_{m_0n_0},\lambda_{m_0n_0}) + \partial_E R_b\delta E_{mn} \nonumber \\& + \partial_\lambda R_b\delta\lambda_{mn}.
\end{align}
Insertion of the lattice coordinates gives
\begin{align}
\label{RB_linear}
&R_b(E_{mn},\lambda_{mn})
\nonumber\\&\simeq R_{B,0}
+
\frac{\pi\hbar}{
[\tilde S_A,\tilde S_B]}
\left(
\delta m[R_b,\tilde S_B]
-\delta n[R_b,\tilde S_A]
\right).
\end{align}
Writing \eqref{plz} as $-\ln P=F(E,\lambda)e^{-2R_b(E,\lambda)/\hbar}$, we evaluate $F=\pi\hbar/(2|\dot\lambda|\,|[\tilde S_A,\tilde S_B]|)$ at the reference crossing and retain the linear variation of $R_B$ from eq. \eqref{RB_linear}. This approximation assumes that the relative variation of $F$ is small over the range of crossings considered here while the changes in $R_b$ of order $\hbar$ change the reflection exponent by order unity. The Landau-Zener probabilities then take the local asymptotic form  
\begin{align}
\label{plz_xpm}
    &P_{mn}\simeq\nonumber\\& \exp\left[-\pi\hbar e^{-\frac{2R_B(E_{m_0n_0},\lambda_{m_0n_0})}{\hbar}}\frac{e^{-2\pi\left[\frac{\delta m[R_B,\tilde{S}_B]-\delta n[R_B,\tilde{S}_A]}{[\tilde{S}_A,\tilde{S}_B]}\right]}}{2|\dot\lambda|\left|\left[\tilde{S}_A,\tilde{S}_B\right]\right|}\right].
\end{align}
Using the reference crossing as the lattice origin, this becomes: 
\begin{align}
\label{P_mn}
&P_{mn}\simeq
\exp\!\left[
-Z\exp(mX_B+nX_A)
\right],\\
&Z = \frac{ \pi\hbar e^{-2R_b(E_{00},\lambda_{00})/\hbar}}{2\left|\dot\lambda\right|\left|\left[\tilde{S}_A,\tilde{S}_B\right]\right|}\\\label{QKNH_XB}
& X_B = -2\pi\frac{[R_b,\tilde{S}_B]}{[\tilde{S}_A,\tilde{S}_B]}\\
&X_A = -2\pi\frac{[R_b,\tilde{S}_A]}{[\tilde{S}_B,\tilde{S}_A]}
\label{QKNH_XA}
\end{align}
The asymptotic form of the transition probability is equivalent to the form we showed in \cite{Stabel_2022} where now the over barrier reflection integral $R_b$ replaces the tunneling integral $T_b$ that appeared in the corresponding equation for the case of a time dependent double well potential. 
\subsection{Sector growth rates and quantum-classical correspondence}
The local probability lattice derived above allows us to identify the quantum separatrix and the dynamical sectors entering the QKNH theorem which we established in \cite{Stabel_2022}. 
The quantum separatrix is defined as the curve $(E_q(\lambda),\lambda)$ satisfying $P(E_q(\lambda),\lambda)=e^{-1}$ where $P$ is the diabatic survival probability. Unlike the classical separatrix energy $E_s(\lambda)=\beta^2(\lambda)$, this quantum boundary is sweep-rate dependent. Around it lies the quantum separatrix zone, in which neither diabatic nor adiabatic passage has probability close to unity. Its width depends on probability thresholds used to define these regimes, as discussed in Sec. \ref{sec:separatrix_zone}. Above the quantum separatrix zone, the avoided crossings are traversed almost completely diabatically. In this regime the dynamics effectively follow the uncoupled semiclassical ladders $E_m^A(\lambda)$ or $E_n^B(\lambda)$, preserving the corresponding rotational branches. Following a diabatic ladder through a crossing means changing between the two instantaneous adiabatic energy levels. We call the states associated with these two kinds of evolution the diabatic quantum sectors $A$ and $B$. Below the quantum separatrix zone, on the other hand, the avoided crossings are traversed adiabatically and the dynamics follow the instantaneous eigenstates which change their rotational character across each crossing. In the diabatic crossing network, this corresponds to a zigzag path alternating between the $A$ and $B$ ladders after every crossing. These adiabatically evolving states define sector $C$. The quantum separatrix therefore marks the boundary between these two dynamical regimes of approximately non-adiabatic and approximately adiabatic evolution rather than the classical transition between rotational and librational motion. Figure \ref{fig:schematic} illustrates the redistribution of an incoming ensemble in sector A between the outgoing sectors $B,C$.

From the asymptotic probability lattice \eqref{P_mn}, the condition defining the quantum separatrix becomes
\begin{equation}
    \ln Z + m_{q}X_B + n_{q}X_A = 0
\end{equation}
Here $m_{q}, n_{q}$ are continuous boundary indices. Their derivatives with respect to $\lambda$ give the rates at which semiclassical states enter or leave the corresponding rotational sectors. Combining this boundary condition with the crossing lattice definition \eqref{lambda_mn} yields the sector growth rates defined in \cite{Stabel_2022}:
\begin{equation}
\frac{dD_A}{d\lambda}=-\Gamma X_A,\quad
\frac{dD_B}{d\lambda}=\Gamma X_B,\quad
\frac{dD_C}{d\lambda}=\Gamma(X_A-X_B),
\end{equation}
where the common factor $\Gamma$ is derived in appendix \ref{sec:app:growth_rates}. We consider $\Gamma>0$ and $X_A>X_B>0$ so that the sector $A$ shrinks while the sectors $B,C$ grow. The semiclassical interpretation in terms of corresponding phase space areas we explained in \cite{Stabel_2022} follows directly from the rotational quantization conditions which count one state per phase-space area $2\pi\hbar$. Evaluated at the quantum separatrix, these determine the continuous boundary indices $m_q(\lambda)=( \tilde{S}_A(E_q(\lambda),\lambda)-\tilde{S}_A(E_{00},\lambda_{00}))/(\pi\hbar)$ and $n_q(\lambda)=(S_B(E_q(\lambda),\lambda)-\tilde{S}_B(E_{00},\lambda_{00}))/(\pi\hbar)$. With the area definitions \eqref{S_AB}, \eqref{tilde_SAB} the smooth state counts satisfy
\begin{equation}
    D_r(\lambda)\simeq
D_{r,0}+\frac{
\Sigma_r(E_q(\lambda),\lambda)
+\hbar\Phi(E_q(\lambda),\lambda)
-C_r}{2\pi\hbar}
\end{equation}
where $r\in\{A,B\}$ and $D_{r,0}$ is a fixed counting offset. Consequently,
\begin{equation}
\label{area_growthrates}
\frac{dD_r}{d\lambda}\simeq
\frac{1}{2\pi\hbar}
\frac{d}{d\lambda}
\left[
\Sigma_r(E_q(\lambda),\lambda)
+\hbar\Phi(E_q(\lambda),\lambda)
\right].
\end{equation}
Thus the rotational sector growth rates are the phase-corrected area growth rates along the local quantum separatrix contour of eq. \eqref{P_mn}, measured in units of $2\pi\hbar$. This is the semiclassical state-counting interpretation discussed in \cite{Stabel_2022}. For numerical estimates, we evaluate the local lattice parameters at the quantum boundary energy $E_q(\lambda)$ obtained from eq. \eqref{boundary_cond}, which generally differs from the classical separatrix energy.
\subsection{Review of the QKNH theorem}
\begin{figure}
    \centering
    \includegraphics[width=\linewidth]{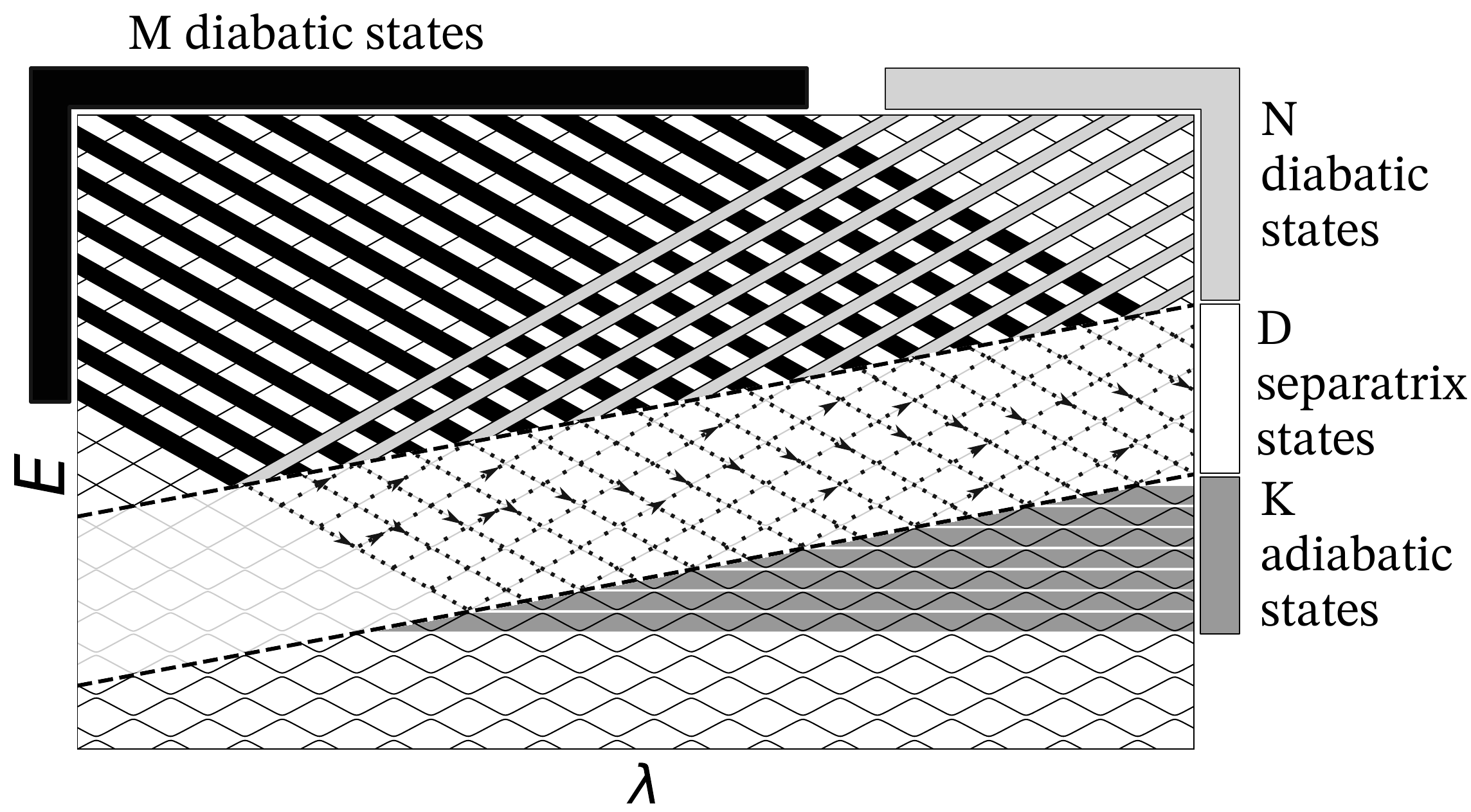}
    \caption{Illustration of an initial ensemble of M diabatic states corresponding to the ladder of energies $\{E_m^A\}$ evolving into the quantum separatrix zone where the probability redistributes over K accessible adiabatic states of sector $C$ on the other side of the separatrix zone and $N$ diabatic accessible states that correspond to the ladder of energies $\{E_n^B\}$.}
    \label{fig:schematic}
\end{figure}
Consider an initial ensemble of $M$ equally weighted consecutive diabatic states $\{\ket{E^A_{0,j}}\}_{j=1}^{M}$ in the shrinking sector $A$,
\begin{equation}
   \rho_0=\frac1M\sum_{j=1}^{M}
\ket{E^A_{0,j}}\bra{E^A_{0,j}}. 
\end{equation}
These states evolve towards the separatrix zone with increasing $\lambda$ (black falling lines in Fig.~\ref{fig:schematic}) and must move entirely into it over time. Let $\lambda_\mathrm{in}$ denote the point at which the last incoming state enters the zone. The ensemble at $\lambda_\mathrm{in}$ is supported on a subspace spanned by N accessible states in sector $B$, K accessible states in sector $C$, and $D$ states in the separatrix zone.  Here, $D$ is an integer state count. The smooth level-count width derived in Sec.~\ref{sec:separatrix_zone} is an approximation of this quantity. In the local lattice approximation, the sector growth rates yield 
\begin{equation}
\label{N_K_estimates}
    N\simeq M\frac{X_B}{X_A},
    \qquad
    K\simeq M\left(1-\frac{X_B}{X_A}\right),
\end{equation}
up to corrections from the finite zone width and discrete level counting \cite{Stabel_2022}. Unitarity preserves the $M$ eigenvalues $1/M$ of the ensemble density matrix. The probability in any $d-$dimensional subspace is thus at most $d/M$. At $\lambda_{\mathrm{in}}$ this bounds the accessible parts of the sectors $B$, $C$ and the zone by $N/M$, $K/M$ and $D/M$, respectively. After passage through the separatrix zone, the final ensemble density matrix is 
\begin{equation}
    \rho_f=U\rho_0U^\dagger,
\end{equation}
where $U$ is the unitary time evolution operator generated by the time-dependent Hamiltonian,
and the population in a final sector $r$ is
\begin{equation}
   \mathcal P_{A\rightarrow r}
=\operatorname{Tr}(\Pi_r\rho_f),
\qquad r=B,C, 
\end{equation}
where $\Pi_r$ projects onto that sector. Outside the separatrix zone, outgoing probability moves away from it. Consequently, after $\lambda_{\mathrm{in}}$, only the probability remaining in the zone can subsequently add to the outgoing sector populations. The bounds above therefore give $\mathcal P_{A\rightarrow B}\le (N+D)/M$ and $\mathcal P_{A\rightarrow C}\le (K+D)/M$. Since total probability is conserved, the two final sector probabilities need to sum to one. The upper bound for each sector probability also gives a lower bound for the other, \textit{i.e.} $P_{A\to B} = 1-P_{A\to C}\geq1-(K+D)/M$ and $P_{A\to C} = 1-P_{A\to B}\geq1-(N+D)/M$. Inserting eq. \eqref{N_K_estimates}, the upper and lower bounds agree up to corrections from the finite zone width and discrete level counting. This yields the weak QKNH result of Ref.~\cite{Stabel_2022}:
\begin{align}
    &\mathcal P_{A\rightarrow B}=
\frac{dD_B/d\lambda}{-dD_A/d\lambda}+ \mathcal{O}(D/M)
=\frac{X_B}{X_A} + \mathcal{O}(D/M)\\
&\mathcal P_{A\rightarrow C}=
1-\frac{X_B}{X_A}+O(D/M).
\end{align}
Its application assumes that the local lattice parameters vary little over the crossings encountered by the ensemble. These estimates apply to an initial ensemble and do not need to hold for an individual initial state. For an exactly periodic lattice, the strong form of the theorem gives the growth-rate ratio without the finite-zone correction after averaging over the dynamical phases, provided the equally weighted initial ensemble spans an integer multiple of lattice periods \cite{Stabel_2022}. 

\section{Explicit analytical expressions for the pendulum}
\subsection{Transition probabilities}
In the following, we compute the parameters $X_{A,B}$ and their ratio explicitly for the pendulum. With $\kappa = (E+\beta^2)/(2\beta^2)$, we define $S=\beta\sqrt{\mu}s(\kappa)$, $R_b = \beta\sqrt{\mu}r(\kappa)$ with 
\begin{align}
&s(\kappa)=8\sqrt{\kappa}\,\label{s_kappa}
\mathrm E\!\left(\kappa^{-1}\right)\\\label{r_kappa}
&r(\kappa)=8\left[
\kappa\mathrm K(1-\kappa)-\mathrm E(1-\kappa)
\right]\\\label{dsdkappa}
& s'(\kappa) = \frac{4}{\sqrt{\kappa}}\mathrm K\left(\kappa^{-1}\right)\\\label{drdkappa}
& r'(\kappa) = 4 \mathrm K\left(1-\kappa\right)
\end{align}
where again $\mathrm{K}$ and $\mathrm{E}$ denote the complete elliptic integrals of the first and second kind. To evaluate the brackets in $X_{A,B}$, we use $[F,G]_{E,\lambda}
=\frac1{2\beta^2}[F,G]_{\kappa,\lambda}$ where the $\lambda-$derivative is taken at fixed $E$ in $[F,G]_{E,\lambda}$ and at fixed $\kappa$ in $[F,G]_{\kappa,\lambda}$. Since the phase correction $\Phi$ is a function of $R_b/\hbar$, we have $[R_b,\Phi]_{E,\lambda}=0$ and hence $[R_b,\tilde{S}]=[R_b,S]$. The brackets in the definitions \eqref{QKNH_XB},\eqref{QKNH_XA} expand to 
\begin{equation}
    [R_b,\tilde{S}_{A,B}]  =-\frac12[R_b,S]\mp\pi[R_b,\alpha]
\end{equation}
and 
\begin{align}
\label{SAB_bracket}
[\tilde S_A,\tilde S_B]_{E,\lambda}
&=-\pi\alpha'\partial_E\tilde S.
\end{align}
With 
\begin{equation}
    \tilde S
    =
    \beta\sqrt{\mu}\,s(\kappa)-\hbar\Phi(x),
\end{equation}
and the phase correction
\begin{equation}
\label{Phi_Imlog}
    \Phi(x)
    =
    \operatorname{Im}\log\Gamma\left(\frac12+ix\right)
    -x\ln x+x,\quad
    x=\frac{\beta\sqrt{\mu}\;r(\kappa)}{\pi\hbar}, 
\end{equation}
we obtain, at fixed $\lambda$,
\begin{equation}
\label{dES}
    \partial_E\tilde S = \frac{1}{2\beta^2}\partial_\kappa\tilde S
    =
    \frac{\sqrt{\mu}}{2\beta}
    \left[
        s'(\kappa)
        -\frac{r'(\kappa)}{\pi}\Phi_x(x)
    \right].
\end{equation}
The derivative of the phase correction is
\begin{equation}
\label{Phi_x}
    \Phi_x(x)= \operatorname{Re}\psi\left(\frac12+ix\right)-\ln x \quad 
\end{equation}
where $\psi(z)=d/dz\ln\Gamma(z)$ denotes the digamma function. We can evaluate the brackets and obtain
\begin{align}
&[R_b,S]_{E,\lambda}
=
\frac{\mu\beta'(\lambda)}{2\beta(\lambda)}
\Big(r'(\kappa)s(\kappa)-s'(\kappa)r(\kappa)\Big) = \frac{8\pi\mu\beta'(\lambda)}{\beta(\lambda)}\\
&[R_b,\alpha]_{E,\lambda}
=\frac{\sqrt{\mu}\,\alpha'(\lambda)}{2\beta(\lambda)}\,r'(\kappa).
\end{align}
where we applied Legendre's identity (see appendix \ref{app:elliptic_integrals}) to obtain $r's-s'r = 16\pi$. Together, these yield 
\begin{align}
X_{A,B}&=
\frac{\pi\sqrt{\mu}}
{\beta\,\partial_E\tilde S}
\left(
r'(\kappa)\pm\frac{8\sqrt{\mu}\,\beta'(\lambda)}{{\alpha'(\lambda)}}
\right)\nonumber\\\label{XAB}
&=\frac{2\pi}{s'(\kappa)-r'(\kappa)\frac{\Phi_x}{\pi}}\left(r'(\kappa)\pm\frac{8\sqrt{\mu}\beta'(\lambda)}{\alpha'(\lambda)}\right)
\end{align}
If we neglect the term involving the phase correction, which is reasonable for energies far above the barrier, we get the simpler expression 
\begin{align}
X_{A,B}&\approx
\frac{2\pi\sqrt\kappa}{\mathrm K(\kappa^{-1})}
\left[
\mathrm K(1-\kappa)
\pm\frac{2\sqrt\mu\,\beta'(\lambda)}{\alpha'(\lambda)}
\right].
\end{align}
For the ratio, the common factor involving the phase correction cancels and we obtain 
\begin{equation}
\label{qknh_ratio}
\frac{X_B}{X_A}
=
\frac{
\alpha'(\lambda)\mathrm K(1-\kappa)-2\sqrt{\mu}\,\beta'(\lambda)
}{
\alpha'(\lambda)\mathrm K(1-\kappa)+2\sqrt{\mu}\,\beta'(\lambda)
}.
\end{equation}
This expression is to be evaluated at the value $\kappa_q(\lambda)$ associated with the quantum separatrix energy $E_q(\lambda)$. 
\subsection{Recovery of the classical KNH probability}
\begin{figure*}[t]
    \centering
    \includegraphics[width=\textwidth]{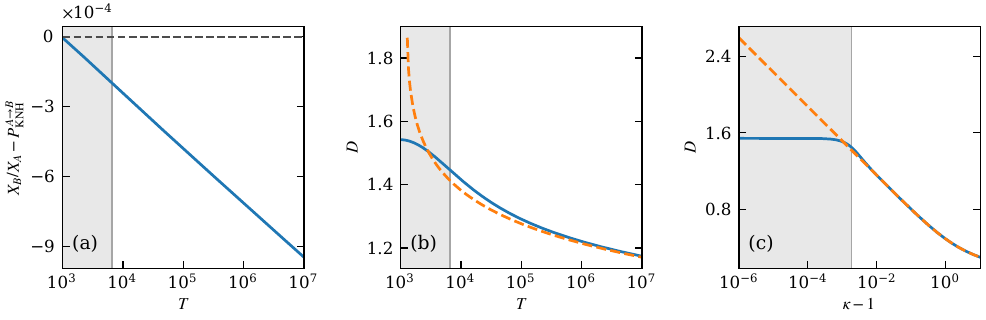}
    \caption{(a) Deviation of the QKNH ratio $X_B/X_A$ from the classical KNH fraction. (b) Width of the separatrix zone over sweep duration T. The boundary parameter $\kappa_q$ has been numerically extracted for 250 logarithmically spaced values in the displayed sweep duration interval before applying the $\kappa$-dependent analytical formula. (c) Width of the separatrix zone according to the analytical formula. In panels (b) and (c), the solid blue curve shows the width formula with the phase correction retained while the dashed orange curve uses the simpler formula that neglects this correction, and diverges near the barrier. Panel (a) uses the numerical boundary obtained with the phase correction retained. In panel (b), the boundary and width are computed separately with (solid blue) and without (dashed orange) the phase correction. The gray shaded region indicates that $R_b/\hbar<1$, outside the weak-reflection regime underlying the derivations of the local lattice parameters and zone width. Inside this shaded region, the formulas are formal extrapolations. Retaining the phase correction does not establish their validity arbitrarily close to the barrier top. For all panels, we used $\mu=50, \hbar=1, \alpha_i= -40.9, \alpha_f=40.9,\beta_i=10,\beta_f=14$, $\lambda = 0.5$, and probability thresholds $P_{\mathrm{high}} = 0.9$, $P_{\mathrm{low}}=0.1$. }
    \label{fig:pendulum_width}
\end{figure*}
In the limit $E\to\beta^2$, \textit{i.e.} $\kappa\to1^+$, we have $\mathrm K(1-\kappa)\to\pi/2$. Hence, we directly obtain from $\eqref{qknh_ratio}$: 
\begin{equation}
  \lim_{\kappa\to1^+}\frac{X_B}{X_A}
=
\frac{\pi\alpha'-4\sqrt{\mu}\,\beta'}
{\pi\alpha'+4\sqrt{\mu}\,\beta'}
=
\mathcal P^{\mathrm{KNH}}_{A\rightarrow B}.
\end{equation}
Thus, as the quantum separatrix energy approaches the classical separatrix energy, we recover the classical KNH probability \eqref{KNH_AB} as expected. Although the ratio has a well-defined algebraic limit as $E\to E_s=\beta^2$, the additional condition $R_b/\hbar\gg1$ requires a bit more care. With $a=\beta\sqrt{\mu}/\hbar$, expansion about $\kappa=1$ gives
\begin{equation}
    \frac{R_B}{\hbar}
    =a\,r(\kappa)
    =2\pi a(\kappa-1)\bigl[1+O(\kappa-1)\bigr].
\end{equation}
The classical separatrix can therefore be approached while maintaining weak reflection in the joint limit
\begin{equation}
    a\to\infty,\qquad
    \kappa-1\to0^+,\qquad
    a(\kappa-1)\to\infty.
\end{equation}
At fixed classical parameters $\beta$, $\mu$ this means letting $\hbar\to 0$ and $E-E_s=2\beta^2(\kappa-1)\to0$ while requiring $a(\kappa-1)\to\infty$. The energy then approaches the classical separatrix energy while $R_B/\hbar\gg1$ remains satisfied. At fixed $\hbar$, however, approaching the barrier top eventually violates the weak-reflection condition.
\subsection{Quantum boundary and separatrix-zone width}
\label{sec:separatrix_zone}
The probability lattice and Landau-Zener formula derived in the preceding sections determine
the location and width of the quantum separatrix zone.
For a linear sweep $\lambda=t/T$, the condition
$P(E_q(\lambda),\lambda)=e^{-1}$ together with the probability formula eqs. \eqref{plz} and \eqref{SAB_bracket} gives
\begin{equation}
\label{boundary_cond}
    \frac{2R_b(E_q(\lambda),\lambda)}{\hbar}
    =
    \ln\left[
        \frac{T\hbar}
        {2|\alpha'\partial_E\tilde S(E_q(\lambda),\lambda)|}
    \right].
\end{equation}
For each value of $\lambda$, this equation can be solved numerically for $E_q(\lambda)$, evaluating both sides of eq. \eqref{boundary_cond} at the trial energy. To estimate the width of the separatrix zone near this boundary, we then use the local probability lattice of eq. \eqref{P_mn}. Within this local approximation, the transition probabilities are given by \eqref{P_mn} with constant $Z,X_A,X_B$. Introducing
\begin{equation}
\label{xi}
    \xi=mX_B+nX_A,
\end{equation}
we have 
\begin{equation}
\label{probability_s}
    P(\xi)=\exp[-Z\exp(\xi)],
\end{equation}
so contours of constant
probability are the straight lines in the $(m,n)$ lattice where $\xi$ is constant. 
We define the transition zone by two probability thresholds,
$0<P_{\mathrm{low}}<P_{\mathrm{high}}<1$.
Inverting the expression \eqref{probability_s} gives
\begin{equation}
    \xi(P)=\ln\left(\frac{-\ln P}{Z}\right).
\end{equation}
and hence the positive threshold interval is
\begin{equation}
    \Delta \xi
    =
    \xi(P_{\mathrm{low}})-\xi(P_{\mathrm{high}})
    =
    \ln\left(
        \frac{-\ln P_{\mathrm{low}}}
             {-\ln P_{\mathrm{high}}}
    \right).
    \label{eq:threshold_interval}
\end{equation}
This interval is independent of $Z$, which translates the transition strip without changing its local width.

To determine how many instantaneous energy levels are within the zone, we have to count states at fixed $\lambda$. Treating the quantum numbers $m$ and $n$ as continuous variables and differentiating the two rotational quantization conditions \eqref{zero_order_quantcond} with respect to energy at fixed $\lambda$ yields 
\begin{equation}
    \frac{d m}{dE} = \frac{\partial_E\tilde{S}_A}{\pi\hbar} = - \frac{\partial_E\tilde{S}}{2\pi\hbar}=\frac{\partial_E\tilde{S}_B}{\pi\hbar}=\frac{d n}{dE}.
\end{equation}
and thus 
\begin{equation}
    \frac{d\xi}{dE}\Big|_\lambda=-\frac{(X_A+X_B)\partial_E\tilde S}{2\pi\hbar}
\end{equation}
Since $\partial_E\tilde{S}>0$ and $X_A+X_B > 0$, $\xi$ decreases with energy and P increases as expected: crossings at higher energies are increasingly diabatic. Since $dm=dn$ and both rotational ladders contribute to the count, \eqref{xi} gives for the smooth level count
\begin{equation}
    D
    \simeq
    |\Delta m|+|\Delta n|
    =
    \frac{2\Delta \xi}{X_A+X_B}.
    \label{eq:level_count_width}
\end{equation}
The actual count is integer-valued and depends on the
positions of the thresholds relative to the discrete
spectrum. 
\begin{figure*}
    \centering
    \includegraphics[width=\linewidth]{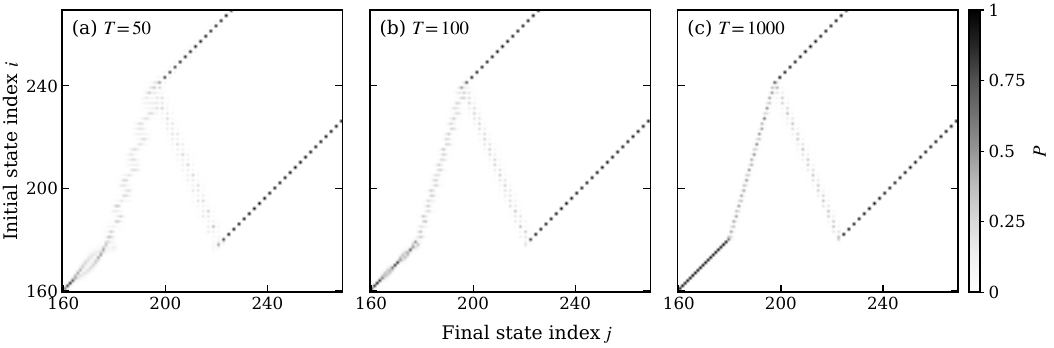}
    \caption{Probabilities of occupying the adiabatic final states $\ket{E_j(T)}$ of the pendulum model after Schrödinger time evolution for a time T with initial states chosen as the eigenstates $\ket{E_i(0)}$ of the instantaneous Hamiltonian at $t=0$. The parameter sweep \eqref{alphasweep}, \eqref{betasweep} is performed with $\alpha_i = -10.9, \alpha_f = 10.9, \beta_i = 10, \beta_f = 11$, $\mu=50$, $\hbar=1$. We use the approximate finite Fourier basis representation in appendix \ref{fourier_hamiltonian} to implement the pendulum Hamiltonian numerically, with a basis cut-off $N=250$, yielding a $(2N+1)\times (2N+1)$ matrix.}
    \label{fig:single_state}
\end{figure*}
Inserting the definitions \eqref{XAB} of the local lattice parameters $X_{A,B}$ into \eqref{eq:level_count_width} yields
\begin{equation}
    D\simeq
    \frac{\Delta \xi}{2\pi}
    \left[
        \frac{s'(\kappa)}{r'(\kappa)}
        -\frac{\Phi_x(x)}{\pi}
    \right],
    \qquad
    x=\frac{\beta\sqrt{\mu}\,r(\kappa)}{\pi\hbar},
    \label{eq:width_phase_corrected}
\end{equation}
and $\Phi_x$ is given by eq. \eqref{Phi_x}. Neglecting the phase correction gives the simpler expression 
\begin{equation}
    D(\kappa)
    \simeq
    \frac{\Delta \xi}{2\pi}
    \frac{\mathrm K(\kappa^{-1})}
         {\mathrm K(1-\kappa^{-1})}.
    \label{eq:width_elliptic}
\end{equation}
This leading width decreases monotonically with $\kappa$. 
For a specified sweep, the relevant width is obtained by evaluating these expressions at $\kappa_q(\lambda)$ associated with the quantum separatrix energy. Increasing the sweep duration $T$ at fixed $\hbar$ shifts the quantum separatrix energy towards higher energies and therefore reduces the leading width. Reducing $\hbar$ at fixed $T$ instead brings the boundary closer to the classical separatrix energy over the parameter range considered, although this trend eventually leaves the weak-reflection regime. Numerical evaluations illustrating both dependencies are shown in Appendix~\ref{sec:app:quantum_boundary}. The corresponding local energy width of the zone is
\begin{equation}
    \Delta E_{D}\simeq \frac{2\pi\hbar\Delta\xi}{(X_A+ X_B)\partial_E\tilde S}
    =
    \frac{\hbar\,\Delta\xi}{2\partial_E R_b} = \frac{\hbar\beta\;\Delta\xi}{4\sqrt{\mu}\;\mathrm{K}(1-\kappa)}.
    \label{eq:energy_width}
\end{equation}
The estimated zone width depends on the threshold convention; below we use $P_{\mathrm{low}}=0.1$ and $P_{\mathrm{high}}=0.9$.


\section{Ensemble splitting: evolution of the quantum pendulum}
\label{sec:transitions}
\begin{figure}
    \centering
    \includegraphics[width=\linewidth]{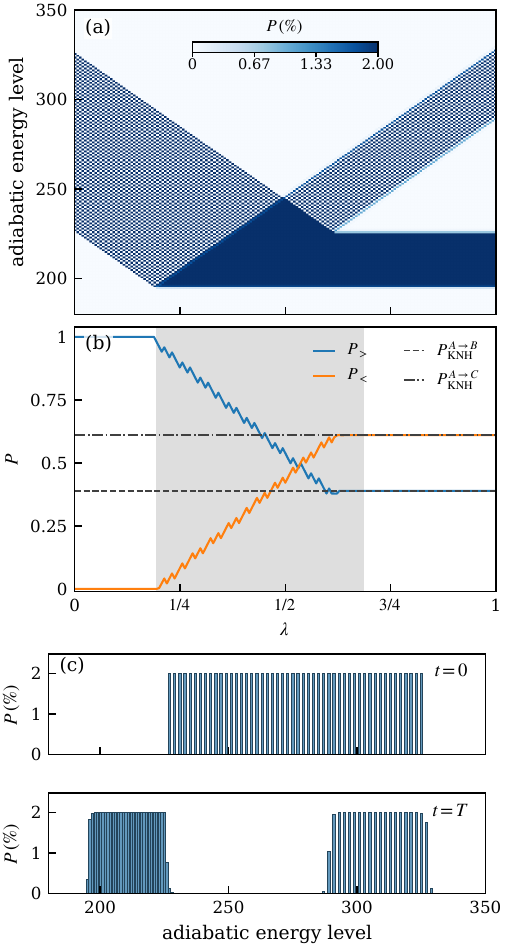}
    \caption{(a) Effective probabilistic evolution of an ensemble of 50 states initially in the shrinking subspace A for the parameter sweep \eqref{alphasweep}, \eqref{betasweep} and $\alpha_i = -40.9$, $\alpha_f= 40.9$, $\beta_i = 10$, $\beta_f = 14$, $\hbar =1$, $\mu=50$, and $T=10^5$ and the probability thresholds $P_{\mathrm{low}}=0.1$, $P_{\mathrm{high}}=0.9$. Each vertical slice shows the adiabatic populations before the next avoided-crossing update. The transition probabilities are computed by eq. \eqref{plz} with the phase correction retained as in Table \ref{tab:boundary_params}. The energy of the intersection is approximated by the mean of the two energy levels that meet at the avoided crossing. The instantaneous energies are obtained as the eigenvalues of the approximate Hamiltonian matrix. (b) Total probability above ($P_{>}$) and below ($P_{<}$) the quantum separatrix zone. Initially, $P_{>}$ is the population of subspace A. After the ensemble has traversed the zone, $P_{>}$ and $P_{<}$ give the populations of the growing subspaces B and C, respectively. The gray interval extends from the first state entering the transition zone to the last state leaving it. (c) Distribution over the adiabatic states at the start and end of the parameter sweep protocol. }
    \label{fig:LZ_evolution}
\end{figure}
Having established the quantum sectors and QKNH transition probabilities for the quantum pendulum, we want to illustrate the evolution of the quantum pendulum under the kind of parameter sweeps we considered here. First, we look at the coherent evolution of the quantum pendulum for different initial states under the parameter sweep \eqref{alphasweep}, \eqref{betasweep} to show the dynamical behavior of individual initial eigenstates. We then consider the evolution of ensembles which the QKNH theorem addresses. For these, we use an effective incoherent probabilistic evolution which becomes valid after protocol averaging. Fig.~\ref{fig:single_state} shows the probability of occupying the adiabatic states $\ket{E_j(T)}$ at the end point $t=T$ of a parameter sweep:
\begin{equation}
P(i,j)
=
\left|
\left\langle E_j(T)\middle|\psi_i(T)\right\rangle
\right|^2,
\end{equation}
where the system is initially prepared in the single instantaneous eigenstate
\begin{equation}
    \ket{\psi_i(0)}=\ket{E_i(0)}
\end{equation}
for three different sweep durations and illustrative parameters. The parameters used here differ from those in the probabilistic ensemble evolution in Fig.~\ref{fig:LZ_evolution} and are chosen to visualize the different dynamical regimes more clearly. These transition probabilities are obtained by solving the time-dependent Schrödinger equation for every initial eigenstate of the instantaneous Hamiltonian shown in the figure separately. We use an approximate Fourier basis representation of the pendulum Hamiltonian with a finite basis cut-off (see appendix \ref{fourier_hamiltonian}). The resulting displayed object is essentially the same as the transition elements displayed in Fig.~9,10 and 12 in Ref. \cite{harpermodel} by Quillen \emph{et al.} except that we show the transition probability which is the square of the absolute value of the transition element used in their figures for the Harper model. 

For rapid parameter sweeps, shown in Fig.~\ref{fig:single_state} (a),(b), transitions occur even between neighboring levels below the classical separatrix energy where there are no avoided crossings. This indicates that the sweep is clearly outside the regime assumed by the QKNH theory where we require adiabatic evolution in between the encounters of avoided crossings. For sufficiently slow sweeps, however, the evolution becomes adiabatic everywhere except in the vicinity of the avoided crossings above the classical separatrix. The QKNH description is restricted to this regime. We can identify several qualitatively different types of evolution depending on the initial states. Initial states below the quantum separatrix zone evolve almost perfectly adiabatically and therefore remain on a single instantaneous eigenstate throughout the sweep. Initial states above the quantum separatrix may instead encounter the quantum separatrix zone while evolving through the network of avoided crossings. In this case the probability splits, resulting in a final probability distribution consisting of two well-separated peaks. For another set of initial states, the probability does not spread over several final states because their diabatic trajectories evolve away from the quantum separatrix zone and therefore do not encounter the transition region during the sweep. Thus the alternating pattern of splitting and no splitting visible in Fig.~\ref{fig:single_state} (c) is a direct consequence of initial states corresponding to the $E_m^A$ or the $E_n^B$ ladder of the avoided-crossing network discussed in Sec.~\ref{subsec:QKNH}. For high energy initial states with index $i\gtrsim240$ in Fig.~\ref{fig:single_state}, no splitting occurs because the states corresponding to the $B$-ladder still evolve diabatically away from the quantum separatrix zone and the states corresponding to the $A$ ladder are on a diabatic path towards the quantum separatrix zone but do not reach it during the sweep. 

The evolution shown in these figures is for the individual eigenstates and thus exhibits coherent interference between different transition paths. The QKNH theorem, however, concerns the evolution of quantum ensembles initially supported entirely on the shrinking sector. In the present system, this corresponds to an ensemble occupying a contiguous range of rotational levels in sector $A$. As shown in Ref.~\cite{Stabel_2022,buerkle_qdimer1}, averaging over the dynamical phases accumulated between successive avoided crossings, which is effectively equivalent to averaging over many slightly different sweep protocols, eliminates coherent interference between different transition paths. The resulting protocol-averaged ensemble dynamics are described by an effective probabilistic evolution through the network of avoided crossings, in which occupation probabilities remain unchanged unless there is an avoided crossing and are redistributed at each avoided crossing according to the corresponding Landau-Zener probability. Specifically, at an avoided crossing between instantaneous levels $j$ and $j+1$, their populations are updated according to
\begin{equation}
\begin{pmatrix}
P_j^{\mathrm{out}}\\
P_{j+1}^{\mathrm{out}}
\end{pmatrix}
=
\begin{pmatrix}
1-P_{\mathrm{LZ}}^{j\to j+1} & P_{\mathrm{LZ}}^{j\to j+1}\\
P_{\mathrm{LZ}}^{j\to j+1} & 1-P_{\mathrm{LZ}}^{j\to j+1}
\end{pmatrix}
\begin{pmatrix}
P_j^{\mathrm{in}}\\
P_{j+1}^{\mathrm{in}}
\end{pmatrix},
\end{equation}
while all other populations remain unchanged.
Here $P_{\mathrm{LZ}}^{j\to j+1}$ = $P_{\mathrm{LZ}}^{j+1\to j}$ is the diabatic survival
probability which corresponds to switching between
the two instantaneous adiabatic levels ${j,j+1}$. We apply these updates successively through the crossing network, without retaining relative phases. The crossing energy entering the analytical Landau-Zener formula given by eq. \eqref{plz} is approximated by the mean of the two instantaneous eigenenergies at the avoided crossing, obtained by diagonalizing the truncated Fourier basis Hamiltonian \eqref{eq:fourier_hamiltonian}.  

As an example, we consider an initial ensemble of 50 equally weighted adiabatic states that correspond to consecutive states in the diabatic sector A. Fig.~\ref{fig:LZ_evolution} (a) shows this incoherent Landau-Zener probabilistic evolution of the occupation probabilities over the dimensionless sweep parameter $\lambda = t/T$. During the evolution shown here, the entire ensemble of initial A states reaches the separatrix zone and passes through it during the parameter sweep. Fig. \ref{fig:LZ_evolution} (c) shows the probability distribution over the adiabatic states at the initial and final time. The sum of the probabilities in each of the two final peaks is what the QKNH theorem predicts. To quantify the ensemble splitting during the evolution, we classify occupied states according to whether they lie above or below the quantum separatrix zone, at each value of $\lambda$. We define
\begin{align}
P_> &= \sum_{j\in\mathcal S_>} P_j,\\
P_< &= \sum_{j\in\mathcal S_<} P_j,
\end{align}
where $\mathcal S_>=\{j:P^{\mathrm{LZ}}_{j\to j+1}>P_\mathrm{high}\}$ and $\mathcal S_<=\{j:P^{\mathrm{LZ}}_{j\to j+1}<P_\mathrm{low}\}$ denote the sets of indices of instantaneous eigenstates above and below the quantum separatrix zone based on the probability thresholds we used throughout this paper. $P_j$ is the probability of occupying the instantaneous eigenstate with eigenvalue $E_j$. The remaining probability is defined as $P_\mathrm{zone}$ and because of probability conservation $P_>+P_<+P_{\mathrm{zone}}=1$. The initial ensemble is confined to sector $A$, which lies entirely above the quantum separatrix, so $P_>=1$ and $P_<=0$. During the evolution, probability is gradually transferred from the shrinking sector $A$ into the two growing sectors $B$ and $C$. After the ensemble has completely passed through the separatrix zone, the sector populations settle to their final values. The higher energy peak consists entirely of the population that has entered the diabatic sector $B$, while the lower energy peak corresponds to the population that has entered the adiabatic sector $C$. Hence, we identify $P_>$ with $P_{A\rightarrow B}$ and $P_<$ with $P_{A\rightarrow C}$ at the final points $\lambda$ after the shaded interval in Fig.~\ref{fig:LZ_evolution} (b). Outside this interval, the evolution is effectively deterministic: before entering the transition region the ensemble occupies only the shrinking quantum sector, while after leaving it the probability is completely distributed between the two growing sectors. 
\begin{table}[]
    \centering
    \begin{tabular}{c|c|c|c|c|c|c}
\toprule
$\lambda$ & $\kappa_q$ & $R_B/\hbar$ & $X_A$ & $X_B$ & $X_B/X_A$ & $D$\\
\midrule
0.00 & 1.0053 & 2.3345 & 3.5025 & 1.3593 & 0.3881 & 1.2688\\
0.25 & 1.0049 & 2.3776 & 3.4709 & 1.3472 & 0.3881 & 1.2803\\
0.50 & 1.0045 & 2.4170 & 3.4425 & 1.3363 & 0.3882 & 1.2909\\
0.75 & 1.0043 & 2.4532 & 3.4166 & 1.3263 & 0.3882 & 1.3006\\
1.00 & 1.0040 & 2.4867 & 3.3929 & 1.3172 & 0.3882 & 1.3097\\
\bottomrule
\end{tabular}
    \caption{Numerical evaluation of the quantum boundary energy parameter $\kappa_q$, local lattice parameters $X_{A,B}$ at the quantum boundary and width of the separatrix zone for the thresholds and sweep protocol used in Fig. \ref{fig:LZ_evolution}. The phase correction is retained here. The parameters are computed for several values of $\lambda$ throughout the sweep. The value of the classical KNH fraction for this protocol is $P_{KNH}^{A\to B}\approx0.3886$. }
    \label{tab:boundary_params}
\end{table}

Table~\ref{tab:boundary_params} shows the local lattice parameters $X_{A,B}$ and their ratio, evaluated at the quantum separatrix energy $E_q(\lambda)$. The boundary remains close to the classical separatrix throughout
the sweep, and the estimated smooth zone width is about $1.3$ levels for the chosen thresholds. The final sector populations in Fig.~\ref{fig:LZ_evolution} are close to the local QKNH ratio $X_B/X_A$, which is itself close to the classical KNH fraction for this sweep.

\section{Summary and discussion}
In this article, we have applied the semiclassical framework we introduced in Ref.~\cite{Stabel_2022} to a quantum pendulum whose rotational levels form, under a slow parameter sweep protocol, a network of avoided crossings above the classical separatrix energy. Instead of ordinary barrier tunneling in the double-well system we considered previously, the coupling between the two semiclassical rotational branches arises from above-barrier reflection. Starting from the semiclassical quantization condition, we constructed the local two-state Hamiltonian, the associated Landau–Zener probabilities, and the lattice of avoided crossings. These probabilities have the same local double-exponential dependence on the lattice indices as in Ref.~\cite{Stabel_2022}, with the above-barrier reflection integral replacing the barrier tunneling integral. The pendulum therefore supplies the complementary semiclassical example in which two dynamically distinct regions lie above the classical separatrix energetically and the third lies below it. For the pendulum, the local lattice parameters $X_{A,B}$, which determine the sector growth rates, and their ratio, which determines the sector transition probability, can be expressed explicitly in terms of complete elliptic integrals. At a specified energy, the common phase-dependent factor cancels from $X_B/X_A$, although the phase correction affects the individual coefficients, the zone width, and the location at which the ratio is evaluated. As the boundary energy approaches the classical separatrix energy, the ratio approaches the classical KNH probability which is the expected semiclassical correspondence. Interpreting this correspondence within the weak-reflection limit requires a joint limit in which the relative distance from the barrier vanishes while $R_B/\hbar$ remains large.

The condition $P(E_q,\lambda) = e^{-1}$ yields the quantum separatrix energy and hence the location of the sweep-rate dependent quantum boundary tied to the relevant transition region. Increasing the sweep duration at fixed $\hbar$ moves the boundary away from the classical separatrix energy, towards higher energies, where reflection is weaker and the leading level-count width is smaller. The exponential energy dependence of reflection makes this displacement relatively small even over several orders of magnitude of the sweep duration. For the protocol examined here, the boundary consequently remains close to the classical separatrix, the coefficient ratio differs only slightly from classical KNH, and the threshold-defined zone has a smooth level-count width of about $1.3$ for the parameters considered. The narrow transition zone is explained by an analytical formula for the zone width for the pendulum: away from the barrier the leading width decreases with energy, and increases only logarithmically as the barrier is approached. The phase-corrected width formula formally approaches a plateau near the barrier top, beyond the weak-reflection regime in which the width formula was derived. Consequently, substantial changes in the boundary location produce only modest changes in the number of levels within the zone. 

The effective incoherent Landau–Zener evolution we showed in Fig.~\ref{fig:LZ_evolution} for an example sweep protocol illustrates the splitting of an ensemble that the QKNH theorem describes. Probability is redistributed while the ensemble traverses the quantum separatrix zone and becomes stationary in the outgoing sectors after the passage is complete. The resulting occupations agree closely with the QKNH ratio defined in Ref.~\cite{Stabel_2022} and reviewed here, and, for these parameters, with classical KNH. Schrödinger time evolution of individual states can nevertheless display coherent interference, and sufficiently rapid sweeps also produce transitions outside the avoided-crossing network. 

The semiclassical construction we use assumes locally isolated avoided crossings, approximately adiabatic evolution between them, and a weak-reflection expansion of the local spectrum. Together with the double-well analysis, the present construction shows how the QKNH description applies to dynamical tunneling in general, including both barrier tunneling and above-barrier reflection. 

\section{Outlook}
In this work, we have focused on the total probabilities of entering each of two growing dynamical quantum sectors from a shrinking sector. A complementary case arises when there is only one growing sector, so that all incoming probability leaving the shrinking sectors enters it with certainty. The unit transfer probability does not specify how the population is distributed among the levels of the receiving sector. In particular, if we perform a forward-and-backward parameter sweep, the ensemble can split at the quantum separatrix on the reverse sweep, leaving only a fraction of the population in the original sector at the end of the cycle. This incomplete return under reversal of the parameter sweep protocol is known as probabilistic hysteresis \cite{buerkle_qdimer1,buerkle_qdimer2}. In a forthcoming article \cite{separatrix_entropy}, we investigate the level populations using a continuum description of the phase-averaged (effectively incoherent) crossing dynamics. For an equally populated ensemble confined to one of the two donor sectors, we show that quantum separatrix crossing produces an effectively uniform distribution over the larger number of accessible states of the single receiving sector. We determine the reduced occupation probability per outgoing state from the local lattice parameters $X_{A,B}$ and derive the resulting increase in diagonal entropy in the instantaneous energy basis. This relates the local lattice parameters and the choice of initially populated sectors to reversibility under forward-and-backward parameter sweeps.

\appendix
\section{Elliptic integrals}
\label{app:elliptic_integrals}
This section supplies the necessary definitions and identities for elliptic integrals that we use in the main text. The complete elliptic integrals of the first and second kind are defined by \cite{gradshteyn2007}:
\begin{align}
&\mathrm K(m)=
\int_0^{\pi/2}
\frac{d\theta}{\sqrt{1-m\sin^2\theta}}\\
&\mathrm E(m)=
\int_0^{\pi/2}
\sqrt{1-m\sin^2\theta}\,d\theta,
\end{align}
where the parameter $m$ is in the literature also commonly expressed in terms of the modulus $k$ with $k^2=m$. The following identities hold for the derivatives:
\begin{align}
&\frac{d\mathrm E(m)}{dm}
=\frac{\mathrm E(m)-\mathrm K(m)}{2m}\\
&\frac{d\mathrm K(m)}{dm}
=\frac{\mathrm E(m)-(1-m)\mathrm K(m)}
{2m(1-m)}.
\end{align}
For $m>0$ it also holds
\begin{align}
\label{ellipk_negativem}
&\mathrm K(-m)=\frac1{\sqrt{1+m}}\,
\mathrm K\!\left(\frac{m}{1+m}\right)\\\label{ellipe_negativem}
&\mathrm E(-m)=\sqrt{1+m}\,
\mathrm E\!\left(\frac{m}{1+m}\right).
\end{align}
Legendre's relation is the following identity that involves the complementary parameter:
\begin{equation}
\label{legendre}
\mathrm K(m)\mathrm E(1-m)
+\mathrm E(m)\mathrm K(1-m)
-\mathrm K(m)\mathrm K(1-m)
=\frac{\pi}{2}.
\end{equation}
A combination of these is used to obtain $r's-s'r=16\pi$ in the main text as we show here. First we use 
\eqref{ellipe_negativem},\eqref{ellipk_negativem} with $m = \kappa-1 >0$ (for $\kappa >1)$ to obtain 
\begin{align}
&\mathrm{K}(1-\kappa) = \frac{1}{\sqrt{\kappa}}\mathrm{K}(1-\kappa^{-1})\\
&\mathrm{E}(1-\kappa) = \sqrt{\kappa}\mathrm{E}(1-\kappa^{-1}).
\end{align}
Then, we use $1-\kappa^{-1}:=\tilde{m}$, and apply Legendre's relation \eqref{legendre}. 
\section{Modified Bohr-Sommerfeld rule}\label{app:quantcond}
The stationary Schrödinger equation for the pendulum model in the coordinate basis can be obtained by promoting the classical canonical coordinates to operators and using $\hat{p} = -i\hbar\partial_\phi$. By shifting the coordinate $\theta = \phi-\pi$, the maximum of the cosine barrier can be shifted to the origin. At fixed $\lambda$, the momentum shift can be eliminated by the transformation $\psi(\theta) = e^{i\theta\alpha(\lambda)/\hbar}\Psi(\theta)$. The resulting time-independent Schrödinger equation for $\Psi(\theta)$ is 
\begin{equation}
    \left[-\frac{\hbar^2}{2\mu}\frac{\partial^2}{\partial\theta^2} + \beta^2(\lambda)\cos(\theta)\right]\Psi(\theta) = E\Psi(\theta)
\end{equation}
For $E>\beta^2$, we define 
\begin{equation}
    P(\theta) = \sqrt{2\mu\left(E-\beta^2\cos\theta \right)}.
\end{equation}
The WKBJ solutions to the right and the left of the barrier top are given by
\begin{equation}
    \Psi_{I,II} = \frac{A_{I,II}}{\sqrt{P(\theta})}e^{\frac{i}{\hbar}\int_0^\theta d\theta'P(\theta')} + \frac{B_{I,II}}{\sqrt{P(\theta})}e^{-\frac{i}{\hbar}\int_0^\theta d\theta'P(\theta')}
\end{equation}
where regions I and II lie to the left and right of the potential maximum, respectively. The angle coordinate is cyclic and the wave function must be single-valued on the ring. Requiring $2\pi$-periodicity of the physical wave function and its derivative,
\begin{equation}
    \psi(\pi)=\psi(-\pi),\qquad
    \psi'(\pi)=\psi'(-\pi),
\end{equation}
then implies the following boundary conditions for the transformed wave function $\Psi$: 
\begin{align}
    \label{eq:twisted_boundary_Psi}
    &\Psi(\pi)=q\Psi(-\pi)\\\label{eq:twisted_boundary_Psi_prime}
    &\Psi'(\pi)=q\Psi'(-\pi),
    \\&q=e^{-2\pi i\alpha/\hbar}.
\end{align}
Since $P(\theta)$ is even,
\begin{equation}
    \int_0^{\pm\pi}P(\theta)\,d\theta=\pm\frac{S}{2}
\end{equation}
with the definition of $S$ in eq. \eqref{S} in the main text. Moreover, $P(\pi)=P(-\pi)$ and $P'(\pi)=P'(-\pi)=0$. Substituting the WKBJ solutions and their derivatives into \eqref{eq:twisted_boundary_Psi} and \eqref{eq:twisted_boundary_Psi_prime} therefore gives
\begin{align}
    A_{\mathrm{II}}e^{iS/(2\hbar)}
    +B_{\mathrm{II}}e^{-iS/(2\hbar)}
    &=
    q\left[
        A_{\mathrm I}e^{-iS/(2\hbar)}
        +B_{\mathrm I}e^{iS/(2\hbar)}
    \right],\\
    A_{\mathrm{II}}e^{iS/(2\hbar)}
    -B_{\mathrm{II}}e^{-iS/(2\hbar)}
    &=
    q\left[
        A_{\mathrm I}e^{-iS/(2\hbar)}
        -B_{\mathrm I}e^{iS/(2\hbar)}
    \right].
\end{align}
Adding and subtracting these equations yields
\begin{equation}
    \begin{pmatrix}
        A_{\mathrm{II}}\\ B_{\mathrm{II}}
    \end{pmatrix}
    =
    q
    \begin{pmatrix}
        e^{-iS/\hbar}&0\\
        0&e^{iS/\hbar}
    \end{pmatrix}
    \begin{pmatrix}
        A_{\mathrm I}\\ B_{\mathrm I}
    \end{pmatrix}.
    \label{eq:periodic_coefficients}
\end{equation}
Above the barrier, two classical rotational branches exist at the same energy. Their semiclassical ladders form avoided crossings because they are coupled by above-barrier reflection. Hence, the coefficients $A_I, B_I$ and $A_{II}, B_{II}$ need to obey the following connection formula \cite{CHILD1974280,child2014semiclassical} in addition to the periodicity condition \eqref{eq:periodic_coefficients}: 
\begin{align}
\label{connectionformula}
    \left(\begin{matrix} A_{II} \\ B_{II}\end{matrix}\right) = \left(\begin{matrix} \sqrt{1+e^{-\frac{2R_b}{\hbar}}}e^{-i\Phi} & -ie^{-\frac{R_b}{\hbar}}\\ ie^{-\frac{R_b}{\hbar}} & \sqrt{1+e^{-\frac{2R_b}{\hbar}}}e^{i\Phi} \end{matrix}\right)\left(\begin{matrix} A_I\\ B_I \end{matrix}\right)\;,
\end{align}
with
\begin{align}\label{Rb_def}
    &R_b = i\int_{\theta_-}^{\theta_+}d\theta P(\theta)\\
    &\Phi = \arg\Gamma\left(\frac{1}{2}+\frac{iR_b}{\pi\hbar}\right) - \frac{R_b}{\pi\hbar}\left(\log\left|\frac{R_b}{\pi\hbar}\right|-1\right)
\end{align}
where $\theta_\pm$ are the complex turning points where $E-\beta^2\cos\theta = 0$ for $E > V_{max} = \beta^2$, and $\arg(z)$ is the phase angle of the complex argument. We use the negative branch of the square root in \eqref{Rb_def}, so $R_b$ is real and non-negative. 

Combining this relation with
\eqref{eq:periodic_coefficients}, a nonzero wave function exists only if
\begin{equation}
    \det
    \begin{pmatrix}
        \sqrt{1+\varepsilon^2}\,e^{-i\Phi}
        -qe^{-iS/\hbar}
        &-i\varepsilon\\
        i\varepsilon
        &\sqrt{1+\varepsilon^2}\,e^{i\Phi}
        -qe^{iS/\hbar}
    \end{pmatrix}
    =0.
\end{equation}
where $\varepsilon = e^{-R_b/\hbar}$.  Expanding the determinant gives
\begin{equation}
    1+q^2
    -2q\sqrt{1+\varepsilon^2}
    \cos\left(\frac{S}{\hbar}-\Phi\right)=0.
\end{equation}
Using
$(q+q^{-1})/2=\cos(2\pi\alpha/\hbar)$,
we obtain the semiclassical quantization condition
\begin{equation}
    \sqrt{1+e^{-2R_B/\hbar}}\,
    \cos\left(\frac{S}{\hbar}-\Phi\right)
    =
    \cos\left(\frac{2\pi\alpha}{\hbar}\right).
    \label{eq:pendulum_quantization_appendix}
\end{equation}
Finally, the definitions \eqref{S_AB}-\eqref{Phase_correction} give $\tilde{S}_A+\tilde{S}_B=-S+\hbar\Phi$ and $\tilde S_A-\tilde S_B=-2\pi\alpha$, yielding the quantization condition \eqref{quandcond} in the main text.  
\section{Numerical implementation of the quantum pendulum model}
\label{fourier_hamiltonian}
We represent the Hamiltonian in the Fourier basis
\begin{equation}
    \varphi_n(\phi) = \braket{\phi|n} = \frac{e^{i n \phi}}{\sqrt{2\pi}}, \qquad n \in \mathbb{Z},
\end{equation}
which forms an orthonormal basis of $L^2[0,2\pi]$. The basis states $\ket{n}$ are eigenstates of the momentum operator $\hat{p}=-i\hbar\partial_\phi$,
\begin{equation}
    \hat{p} \ket{n} = \hbar n \ket{n},
\end{equation}
and satisfy $e^{\pm i\phi} \ket{n} = \ket{n \pm 1}$.

Using $\cos\hat{\phi} = \frac{1}{2}(e^{i\hat{\phi}} + e^{-i\hat{\phi}})$, the matrix elements of the Hamiltonian are
\begin{equation}
    \braket{n' | \hat{H} | n} =
    \frac{(\hbar n - \alpha)^2}{2\mu} \delta_{n,n'}
    - \frac{\beta^2}{2} \left( \delta_{n',n+1} + \delta_{n',n-1} \right).
\end{equation}

The Hamiltonian can thus be written as
\begin{align}
    \hat{H}(\lambda) &= \sum_{n=-\infty}^{\infty} \frac{(\hbar n - \alpha(\lambda))^2}{2\mu} \ket{n}\bra{n}
    \nonumber\\&- \frac{\beta^2(\lambda)}{2} \sum_{n=-\infty}^{\infty} \left( \ket{n}\bra{n+1} + \ket{n}\bra{n-1} \right).\label{eq:fourier_hamiltonian}
\end{align}

For numerical computations, we truncate the basis to $|n| \leq N$, yielding a finite $(2N+1)\times(2N+1)$ matrix representation of $\hat{H}$ and we set $\hbar=1$ if not stated otherwise. Figures \ref{fig:spectrum}, \ref{fig:LZ_heatmap}, and \ref{fig:LZ_evolution} use the basis cut-off $N = 300$, and Fig.~\ref{fig:single_state} uses $N=250$. The indexing of the eigenvalues of this matrix used in Fig.~\ref{fig:single_state} starts from $0$. 
\section{Scaling of the quantum boundary with $\hbar$ and $T$}\label{sec:app:quantum_boundary}
To locate the quantum separatrix energy $E_q(\lambda)$ for the linear sweep $\lambda=t/T$, we solve eq. \eqref{boundary_cond} which follows from \eqref{plz}. Using eq. \eqref{dES}
we obtain, at fixed $\lambda$, 
\begin{align}
    &g_{\Phi}(\kappa;\lambda,T)
    =\nonumber\\
    &2a\,r(\kappa)
    -
    \ln\left[
        \frac{T\hbar\beta(\lambda)}
        {\left|\alpha'(\lambda)\sqrt{\mu}\,
            s'(\kappa)
            -\dfrac{r'(\kappa)}{\pi}
             \Phi_x\left(\dfrac{a\,r(\kappa)}{\pi}\right)
        \right|}
    \right].
    \label{eq:app_g_phase}
\end{align}
where $a = \beta\sqrt{\mu}/\hbar$ and the functions are defined in eqs. \eqref{s_kappa}-\eqref{drdkappa}. We solve $g_\Phi(\kappa;\lambda,T)=0$ for $\kappa_q>1$ and obtain $E_q(\lambda)=\beta^2(\lambda)\,[2\kappa_q(\lambda)-1]$. 
Neglecting the phase correction simplifies the expression by setting $\Phi_x=0$. Repeating the calculation at different $\lambda$ values traces the boundary through the sweep.
\begin{figure}
    \centering
    \includegraphics[width=\linewidth]{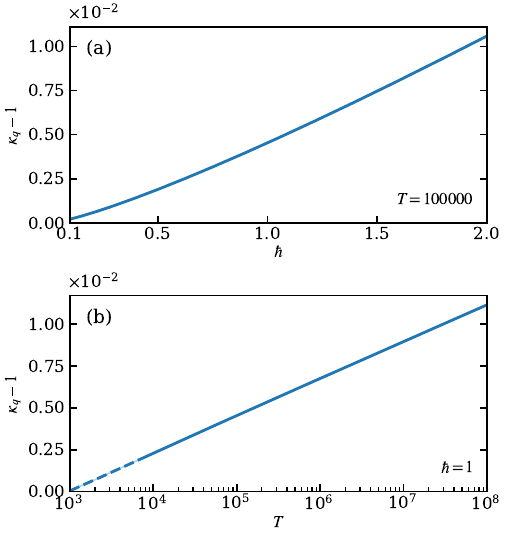}
    \caption{Distance of the quantum boundary from the classical separatrix at $\lambda = 0.5$ and the parameter sweep \eqref{alphasweep}, \eqref{betasweep} with $\alpha_i = -40.9$, $\alpha_f= 40.9$, $\mu=50$, $\beta_i = 10$, $\beta_f = 14$. (a) shows the dependence on $\hbar$ for fixed $T=10^5$, and (b) shows the dependence on $T$ for fixed $\hbar =1$. Each panel uses 200 values of the parameter in the interval shown on the x-axis. Dashed lines indicate that $R_b/\hbar<1$, which is in violation of the weak reflection condition $R_b/\hbar\gg1$ underlying the derivation of the effective Landau-Zener formula. }
    \label{fig:boundary_scaling}
\end{figure}
Figure~\ref{fig:boundary_scaling} illustrates the parameter dependence of the quantum separatrix energy. Here, we solve the root condition for the boundary at $\lambda=0.5$ while retaining the phase correction $\Phi$.  The distance from the classical separatrix is expressed as $\kappa_q-1=(E_q-E_s)/(2\beta^2)$. The two panels vary $\hbar$ at fixed $T=10^5$ and $T$ at fixed $\hbar=1$, respectively, while keeping the classical sweep protocol unchanged, and for each we evaluated 200 parameter values. Dashed segments indicate $R_B/\hbar<1$, where the
weak-reflection condition is violated.
\section{Growth rates of the quantum sectors}
\label{sec:app:growth_rates}
We compute the sector growth rates within the local lattice approximation of eq. \eqref{P_mn}, in which $Z,X_{A},X_B$ are held fixed, in analogy to Ref. \cite{Stabel_2022}. Let $m_q(\lambda), n_q(\lambda)$ be the continuous lattice indices along the quantum separatrix contour $E_q(\lambda)$ for which $P=e^{-1}$. They satisfy
\begin{equation}
    \ln Z+m_q(\lambda)X_B+n_q(\lambda)X_A=0,
\end{equation}
Differentiating with respect to $\lambda$ gives
\begin{equation}
    X_B\frac{dm_q}{d\lambda}
    +X_A\frac{dn_q}{d\lambda}=0.
    \label{eq:app_boundary_index_rates}
\end{equation}
The derivatives therefore have the form
\begin{equation}
    \frac{dm_q}{d\lambda}=-\Gamma X_A,
    \qquad
    \frac{dn_q}{d\lambda}=\Gamma X_B,
    \label{eq:app_index_rates}
\end{equation}
with a common proportionality factor $\Gamma$.
To determine this factor, we use the crossing-parameter map derived in the main text:
\begin{equation}
    d\lambda
    =
    \frac{\pi\hbar}
    {[\tilde S_A,\tilde S_B]_{E,\lambda}}
    \left(
        \partial_E\tilde S_A\,dn_q
        -\partial_E\tilde S_B\,dm_q
    \right).
\end{equation}
Substitution of Eq.~\eqref{eq:app_index_rates} and division by $d\lambda$ yields
\begin{equation}
    1=
    \frac{\pi\hbar\Gamma}
    {[\tilde S_A,\tilde S_B]_{E,\lambda}}
    \left(
        X_B\partial_E\tilde S_A
        +X_A\partial_E\tilde S_B
    \right).
\end{equation}
Hence
\begin{equation}
    \Gamma
    =
    \frac{[\tilde S_A,\tilde S_B]_{E,\lambda}}
    {\pi\hbar
    \left(
        X_B\partial_E\tilde S_A
        +X_A\partial_E\tilde S_B
    \right)}.
    \label{eq:app_Gamma}
\end{equation}
which is the factor appearing in the subspace growth rates of Ref.~\cite{Stabel_2022}. Hence ~\eqref{eq:app_index_rates} gives
\begin{equation}
    \frac{dD_A}{d\lambda}=-\Gamma X_A,
    \qquad
    \frac{dD_B}{d\lambda}=\Gamma X_B.
\end{equation}
Within the local approximation, the dimension of the
separatrix zone is constant. Conservation of the total
number of states then implies
\begin{equation}
    \frac{dD_C}{d\lambda}
    =
    -\frac{dD_A}{d\lambda}
    -\frac{dD_B}{d\lambda}
    =
    \Gamma(X_A-X_B).
\end{equation}
For $\Gamma>0$ and $X_A>X_B>0$, sector $A$ shrinks
while sectors $B$ and $C$ grow. For the pendulum, Eq.~\eqref{eq:app_Gamma} simplifies.
Using
\begin{equation}
    \partial_E\tilde S_A
    =
    \partial_E\tilde S_B
    =
    -\frac12\partial_E\tilde S,
    \qquad
    [\tilde S_A,\tilde S_B]_{E,\lambda}
    =
    -\pi\alpha'\partial_E\tilde S,
\end{equation}
we obtain
\begin{equation}
    \Gamma
    =
    \frac{2\alpha'}{\hbar(X_A+X_B)}.
    \label{eq:app_Gamma_pendulum}
\end{equation}
The common phase-dependent energy derivative cancels
in this relation; no neglect of the phase correction
is required.

\bibliography{bibliography}

\end{document}